%% file: subdimensional_criticality.tex
\documentclass[prb,twocolumn,aps,groupedaddress,nofootinbib]{revtex4}
\input{subdimensional_criticality_preamble.tex}

\usepackage{dcolumn}
\newcommand{\ex}{excitation } 
\newcommand{\exs}{excitations }

\newcommand{\lij}{{\lan ij \ran}}
\newcommand{\zfo}{\mathbb{Z}_4}

\newcommand{\lijd}{{\langle \sfi\sfj\rangle}}
\renewcommand{\ij}{\lij}

\begin{document}

	\title{Subdimensional criticality: condensation of lineons and planons in the X-cube model}
	
	\author{Ethan Lake}
	\affiliation{Department of Physics, Massachusetts Institute of Technology, Cambridge, MA, 02139, USA}
	\author{Michael Hermele} 
	\affiliation{Department of Physics and Center for Theory of Quantum Matter, University of Colorado, Boulder, CO, 80309, USA} 
	
	\begin{abstract}
		
	We study quantum phase transitions out of the fracton ordered phase of the $\zn$ X-cube model. These phase transitions occur when various types of sub-dimensional excitations and their composites are condensed. The condensed phases are either trivial paramagnets, or are built from stacks of $d=2$ or $d=3$ deconfined gauge theories, where $d$ is the spatial dimension. {The nature of the phase transitions depends on the excitations being condensed.	Upon condensing dipolar bound states of fractons or lineons, for $N \geq 4$ we find stable critical points described by decoupled stacks of $d=2$ conformal field theories. Upon condensing lineon excitations, when $N > 4$ we find a gapless phase intermediate between the X-cube and condensed phases, described as an array of $d=1$ conformal field theories. In all these cases, effective subsystem symmetries arise from the mobility constraints on the excitations of the X-cube phase and play an important role in the analysis of the phase transitions.}
	\end{abstract}
	
	\maketitle

\section{Introduction and summary \label{sec:intro}} 

Fracton phases\cite{chamon2005quantum,haah2011local,yoshida2013exotic,vijay2016fracton,pretko2017subdimensional,nandkishore2019fractons,pretko2020fracton} are an exciting class of states of matter in three dimensions which defy many expectations for the types of behavior one expects to see in zero-temperature quantum systems. They have subextensively large ground state degeneracy, excitations whose motions are restricted to lie along low-dimensional submanifolds of space, and provide examples of gapped phases not described within the context of conventional topological quantum field theory.\footnote{We will not attempt a review of the many interesting recent advances in the field of fractons; for an introduction to the literature the reader may consult the reviews\cite{nandkishore2019fractons,pretko2020fracton} and the references therein.}

Fracton phases can be separated into two types, depending on the nature of their excitations.\cite{vijay2016fracton} 
Type-I fracton phases possess mobile excitations, which are typically \emph{lineons} or \emph{planons} that can move along one- or two-dimensional submanifolds of space, respectively.
One paradigmatic example of a type-I fracton phase is the X-Cube (XC)  model.\cite{vijay2016fracton} Type-II phases by contrast do not have any non-trivial mobile excitations, with Haah's code\cite{haah2011local} being the classic example. 

An important question in fracton physics is where fracton phases may arise in the phase diagrams of various systems. One way of approaching this problem is to examine what sorts of continuous quantum phase transitions can occur between a phase with fracton order and a more conventional phase of matter. This question has already been partially addressed in the literature. Previous works have identified first-order quantum phase transitions out of fracton ordered phases via a range of techniques,\cite{devakul2018correlation,slagle2017fracton,muhlhauser2020quantum} and studied the breakdown of fracton order in series expansions.\cite{poon2020quantum} Ref.~\onlinecite{vijay2017isotropic} exploited a duality to 3d $\zn$ gauge theory to propose a continuous transition between the $\zn$ XC phase and a trivial gapped phase; we comment on this proposal in Sec.~\ref{sec:disc}. {(Here and throughout the paper, we refer to spatial dimension rather than space-time dimension, unless explicitly stated otherwise.)} A recent work introduced the notion of hybrid fracton orders, which combine the fully mobile excitations of 3d topological orders with the restricted-mobility excitations of fracton phases, and found continuous phase transitions between such hybrid phases and fracton-ordered phases.\cite{tantivasadakarn2021hybrid} However, fracton order is present on both sides of these phase transitions.
 Other recent works have studied continuous quantum phase transitions in two-dimensional systems with subsystem symmetry but without fracton order.\cite{you2020fracton,you2021fractonic,zhou2021fractal}

In this work, our focus will be on continuous phase transitions out of fracton orders, where the continuous nature and stability of the phase transitions may be demonstrated rigorously. {We will study both critical points, which can be accessed by tuning a single parameter, as well as stable gapless phases proximate to fracton orders.}
Given that the excitations in a fracton phase have restricted mobility, it is natural to imagine that any putative critical point would necessarily involve a rather unconventional field theory description, perhaps of a similar flavor to the novel types of field theories that have appeared when analyzing fracton phases themselves.\cite{slagle2017quantum,slagle2019foliated,you2020fractonic,seiberg2020exoticI,seiberg2020exoticII,seiberg2020exoticIII,seiberg2020moreexotic,slagle2021foliated,hsin2021comments}

The easiest place to start when thinking about continuous phase transitions out of fracton phases is to examine phase transitions in type-I fracton phases where certain types of excitations with restricted mobility are condensed. In these types of phase transitions, it is however not clear whether or not one should expect a critical point characterized by scale invariance, and to what extent the standard theory of critical phenomena should be applicable. 

In this paper we will make some first steps towards addressing these questions. We will focus on what appear to be the simplest examples of continuous phase transitions out of fracton orders, viz. various types of condensation transitions in the 
$\zn$ X-Cube (XC) model.\cite{vijay2016fracton,vijay2017isotropic} {The nature of the condensed phase varies depending on which type of excitation is condensed. In some of the phase transitions, the condensed phase is trivial, while in others it possesses topological or fracton order (distinct from that of the $\zn$ X-cube phase).}

\begin{figure}
	\includegraphics[width=.45\textwidth]{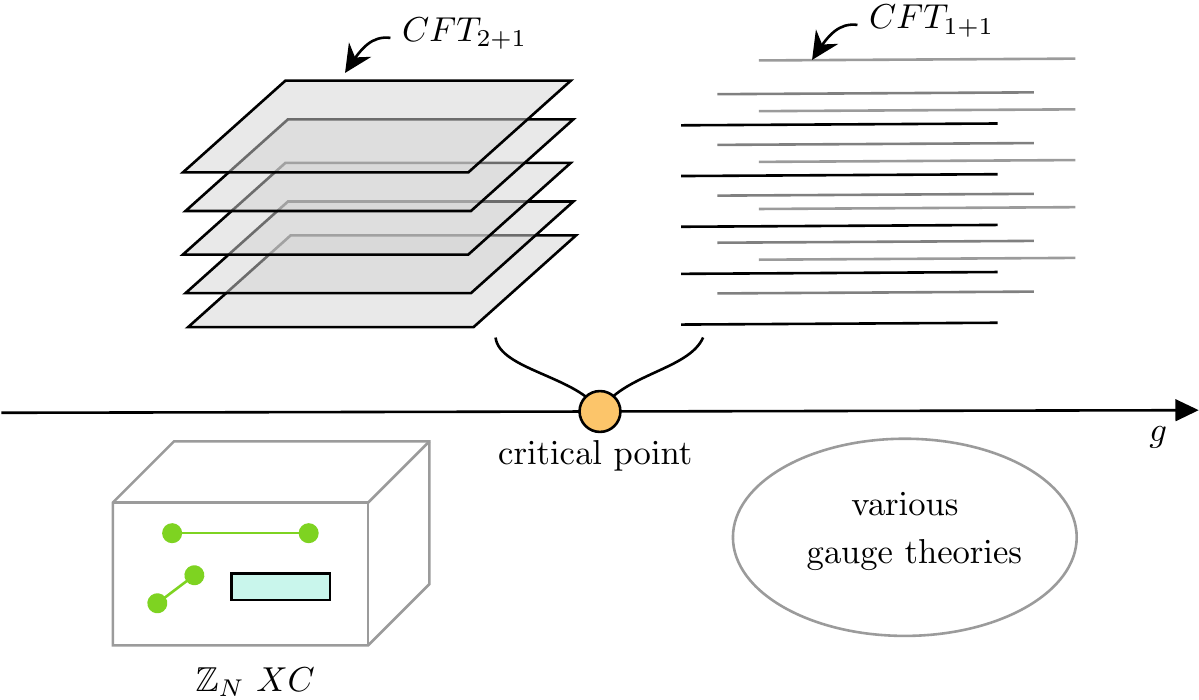}
	\caption{\label{fig:inspirational_schematic} A schematic of the phase transitions we consider. 
	Tuning a parameter $g$ drives a quantum phase transition between the $\zn$ X-Cube model and {a condensate of non-trivial lineon or planon excitations. The resulting condensed phase is either trivial or is some form of deconfined gauge theory. A critical point or intermediate gapless phase can be described as a stack of 2d or 1d CFTs.}} 
\end{figure}

Taking inspiration from the many approaches which construct fracton phases by coupling various well-understood lower-dimensional phases together,\cite{ma2017fracton,vijay2017isotropic,halasz2017fracton,prem2019cage,aasen2020topological,wen2020systematic,sullivan2020fracton,slagle2019foliated} our approach to studying phase transitions in these models will be to look for critical points, or intermediate gapless phases, which can be understood within the context of 1d or 2d critical phenomena.  These critical points and gapless phases arise when condensing planon or lineon excitations of the $\zn$ XC model, and can be described as stacks / arrays of familiar 2d / 1d conformal field theories (see figure \ref{fig:inspirational_schematic} for a schematic). 	

An interesting feature of the critical points and gapless phases we study is the emergence of effective subsystem symmetries. By definition, subsystem symmetries are those which act on degrees of freedom lying only in a subspace, which may be a line, a plane, or even a fractal. Subsystem symmetries are interesting in their own right and also play a variety of important roles in the theory of fracton phases. The effective subsystem symmetries in the theories we study arise from the gauge structure of the $\zn$ X-cube phase, and are robust to arbitrary perturbations, even though such symmetries are not assumed to be present microscopically.

	\begin{table}
		\renewcommand{\arraystretch}{1.4}
		\begin{tabular}{@{}lll@{}}  
			
			sector\,\, & condensate\,\, & condensed phase  \\ \hline 
			
			lineon & $\mfe^z$ & $\zz_{N,2}$ stack   \\ 
			& $\mfe^{y,z}$ & trivial \\  
			& $\mfd^z_\mfe$ & two $\zz_{N,2}$ stacks \\ 
			&  $\mfd^{y,z}_\mfe$ & $\zz_{N,3}$ + $\zz_{N,2}$ stack \\ 
			&  $\mfd^{x,y,z}_\mfe$ & $\zz_{N,3}^2$\\ \hline 
			fracton\,\, & $\mfd^z_\mfm$ & 
			anisotropic model  \\ 
			& $\mfd^{y,z}_\mfm$ & $\zz_{N,2}$ stack\\ 
			& $\mfd^{x,y,z}_\mfm$ & $\zz_{N,3}$

		\end{tabular}
		\caption{\label{tab:condensed_phases} The phases obtained when condensing various types of \exs in the $\zn$ XC model. $\mfe^a $ denotes a lineon mobile in the $a$-direction, while 
		$\mfd^a_\mfe$ and $\mfd^a_\mfm$ are dipolar bound states of lineons and fractons, respectively, with dipole moment in the $a$-direction; see Sec.~\ref{sec:preliminaries} for more details. $\zz_{N,d}$ denotes a deconfined $\zn$ gauge theory in $d$ spatial dimensions, a ``stack'' refers to a decoupled set of theories, and ``trivial'' refers to a paramagnet lacking any topological order. We have identified continuous phase transitions involving lineon condensation provided that $N>4$, and continous transitions involving dipole condensation provided that $N>3$. ``Anisotropic model'' refers to the model of lineons and planeons discussed in Ref. \onlinecite{shirley2019fractional}.}
	\end{table}
	
A summary of this paper is as follows. We start in section \ref{sec:preliminaries} by reviewing some facts about the $\zn$ XC model which will be relevant for the following sections. In section \ref{sec:frac_dip} we study what happens when dipoles of fractons, which are planons, are condensed. In order to identify the resulting condensed phases we employ a dual representation of the XC model in terms of a generalized gauge theory, which is detailed in section \ref{sec:frac_dip_phases}. Condensing fracton dipoles allows single fractons, which in the XC model are immobile, to move freely in directions parallel to the moments of the condensed dipoles. When dipoles with a single direction of dipole moment are condensed the resulting phase is an anisotropic model studied in Ref. \onlinecite{shirley2019fractional}, when two orthogonal directions of dipoles are condensed one finds a stack of deconfined 2d $\zn$ gauge theories, and when all dipoles are condensed the resulting phase is a deconfined $\zn$ gauge theory, with single fractons serving as the gauge charges. In section \ref{sec:2dphase_transitions_fracton} we examine the nature of the condensation transition. When $N<4$ the condensation transitions are {likely to be} generically first order, while when $N\geq 4$ continuous phase transitions exist, and are described by stacks of critical 2d XY models.

Section \ref{sec:2d_lineons} is analogous to the previous section, except we instead consider the condensation of lineon dipoles. When lineon dipoles condense the mobility restrictions on single lineons are relaxed, and single lineons become capable of moving freely in two or three dimensions. When a single orientation of lineon dipole condenses the resulting phase is two interpenetrating stacks of 2d $\zn$ gauge theories, when two orientations of dipoles condense one obtains a 3d $\zn$ gauge theory and a stack of 2d $\zn$ gauge theories, and when all dipoles condense the result is a 3d $\zn^2$ gauge theory. These results are summarized in table \ref{tab:condensed_phases}, and are obtained by utilizing a generalized gauge theory which makes the nature of the condensed phases very explicit, and which is explained in detail in section \ref{sec:phases_lineon_dipoles}. The condensation phase transitions in this case are exactly the same as those that occur when condensing fracton dipoles. 

In section \ref{sec:1dexps} we discuss the slightly simpler problem of condensing lineons. When only one type of lineon condenses the resulting phase is a stack of 2d $\zn$ gauge theories, while if more condense one obtains a trivial paramagnet. In section \ref{sec:oned_dual_hams} we introduce yet another generalized gauge theory, which is used to obtain a field theory description of lineon condensation.
In sections \ref{sec:oned_critical_smalln} and \ref{sec:oned_critical_largen} we show that for $N\leq 4$ the condensation transition is likely to always be first order, while for $N > 4$ it is possible for an intermediate gapless phase to exist in between the X-Cube and condensed phases. This gapless phase is described by an array of strongly coupled Luttinger liquids, and is related to the 3d sliding phases discussed in Ref.~\onlinecite{mukhopadhyay2001sliding}. {The stability of this gapless phase is somewhat delicate, and we argue that (1) by adding inter-chain derivative couplings the phase can be made stable down to very low temperatures and (2) a truly stable phase with more complicated inter-chain couplings should exist in principle.}

Finally, section \ref{sec:disc} contains a conclusion and a brief discussion of some remaining open questions.

\begin{figure}
	\includegraphics[width=.5\textwidth]{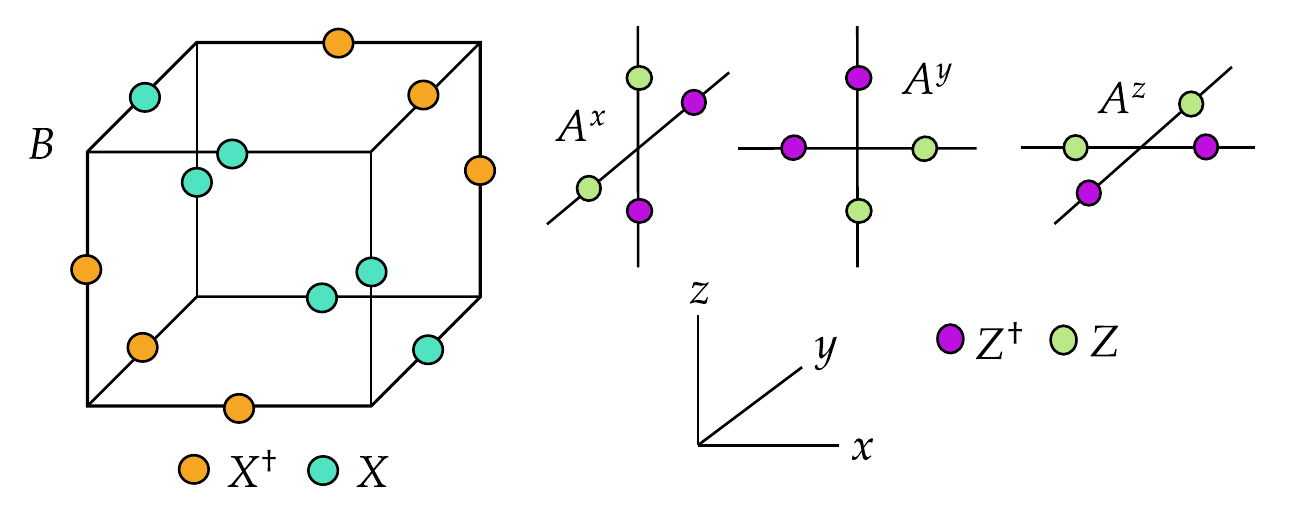}
	\caption{\label{fig:xc_ham} A graphical representation of the terms in the Hamiltonian of the $\zn$ XC model.} 
\end{figure}

\section{Preliminaries \label{sec:preliminaries}}

Before starting in earnest, we will quickly summarize a few important properties of the $\zn$ X-cube (XC) model,\cite{vijay2016fracton,vijay2017isotropic} as well as some of the notation that we will be making frequent use of. 
$Z,X$ will denote $\zn$ clock and shift operators, which obey the relations 
\be ZX = e^{\twp i /N} XZ, \qquad Z^\da X = e^{-\twp i /N} XZ^\da, \ee %
where $N$ will be inferred from context. 
When discussing microscopic Hamiltonians, we will be making use of various dual representations. These dual representations will be in terms of discrete $\zn$ matter spins and discrete $\zn$ gauge fields. The clock and shift operators for the matter fields will always be written in serif font ($\sfz,\sfx$), while those for the gauge fields will always be in mathcal ($\mcz,\mcx$).

The Hilbert space of the $\zn$ XC model is built from $\zn$ spins placed on the links of a 3d cubic lattice. We will write the Hamiltonian for the commuting projector limit of the XC model as 
\be \label{xcham} H_{XC} = -g \sum_i (A_i^x + A_i^y + A^z_i) -K \sum_c B_c + h.c.
\ee 
where the sums over $i$ and $c$ run over the sites and minimal cubes of the lattice, respectively. The operators $A^a_i$ are defined as products of $Z$ operators over the four links touching a given vertex $i$, while the $B_c$ operators are defined as products of $X$ operators over the twelve links of the cube $c$.  Half of the $X$ and $Z$ operators appearing in the $A^a_i$ and $B_c$ terms are Hermitian conjugated in the manner shown in figure \ref{fig:xc_ham}, which ensures that all of the terms in \eqref{xcham} commute, and that 
\be \label{aprodone} A^x_i A^y_i A^z_i = \unit\ee 
at every site $i$. 

Sometimes it is convenient to refer to links of the lattice in an oriented manner. We imagine that each link comes with a given orientation in the $+x$, $+y$ or $+z$ direction. Then, denoting by $\langle i j \rangle$ a pair of nearest-neighbor sites, and letting $\uva$ be the unit vector pointing from $i$ to $j$, we define $X_{\langle i j \rangle} = X$ if $\uva$ is parallel to the link's given orientation, and $X_{\langle i j \rangle} = X^\dagger$ if $\uva$ is anti-parallel to the given orientation. Note that $X_{\langle i j \rangle} = X^\dagger_{\langle j i \rangle}$. The same notation and conventions are used to define $Z_{\lij}$. For simplicity, we will use $l$ to denote links when keeping track of the orientation is not necessary.

The excitations of the Hamiltonian \eqref{xcham} are divided into two sectors. The first sector consists of lineon excitations and their composites. Single lineons correspond to violations of two of the $A^a_i$ terms, and come in three species, which we will denote as $\mfe^x,\mfe^y,\mfe^z$. An $\mfe^x$ excitation at a given site $i$ corresponds to a violation of the two terms $A^y_i$ and $A^z_i$, and similarly for $\mfe^y,\mfe^z$. Our convention is that acting on a ground state with $X_{\langle i j \rangle}$, where \emph{e.g.} $\langle i j \rangle$ is directed along the positive or negative $x$-direction, creates an $\mfe_i^x$ lineon at site $i$ and a $(\mfe_j^x)^\dagger$ antilineon at site $j$.  Each isolated $\mfe^a$ lineon may only move along the $\uva$ direction; any attempts to move it along the directions normal to $\uva$ necessarily result in the creation of additional excitations. String operators for the lineons are formed from products of $X_\lij$ operators in their direction of motion.
Due to the relation \eqref{aprodone}, the lineons obey the fusion rule 
\be \label{fusetoone} \mfe_i^x \mfe^y_i \mfe^z_i = 1.\ee 
Because of this fusion rule, a lineon $\mfe^x$ moving along the $\uvx$ direction may split into an $(\mfe^y)^\da$ antilineon and an $(\mfe^z)^\da$ antilineon, and likewise for the other two species.

{The mobility of individual and multiple lineons can be understood by viewing $\mfe^a$ as carrying a unit $\zn$ charge in each of the two lattice planes that intersect at the $a$-axis along which $\mfe^a$ is free to move. (This was discussed in Ref.~\onlinecite{pai2019fracton} for the $\zt$ X-cube model; the generalization to $\zn$ is straightforward.) The $\zn$ charge in each plane is conserved in the sense that it cannot be changed by any local operator. This gives rise to effective planar $\zn$ subsystem symmetries that will be important in understanding critical points where lineons and their composites condense.}

Another interesting feature of the XC model is the behavior of dipolar\footnote{Since lineons are $\zn$ objects, the ``dipole moment'' of two lineons is rather ill-defined. In what follows we will however abuse terminology and refer to a bound state of two lineons on neighboring vertices as a lineon ``dipole''.} bound states of lineons. Consider a bound state of a pair of $\mfe^b$ and $(\mfe^b)^\dagger$  lineons separated by a given distance in the $\uva$ direction, where $a\neq b$. Because the two lineons cannot move in the $\uva$ direction, they cannot move together and annihilate to vacuum, and indeed it can be shown that an isolated dipole of this sort is stable, in the sense that it cannot be destroyed by any local process.\footnote{By contrast, if lineon and antilineon are separated in the $\uvb$ direction, they can annihilate, and such an excitation is trivial.}
One consequence of the fusion rule \eqref{fusetoone} is that such dipolar bound states may move freely within the plane perpendicular to $\uva$, as they may change the direction of their motion within the plane by having their constituent charges exchange an $\mfe^a$ lineon. However motion parallel to $\uva$ is not possible, and consequently such dipolar bound states are planon excitations.

\begin{figure}
	\includegraphics[width=.48\textwidth]{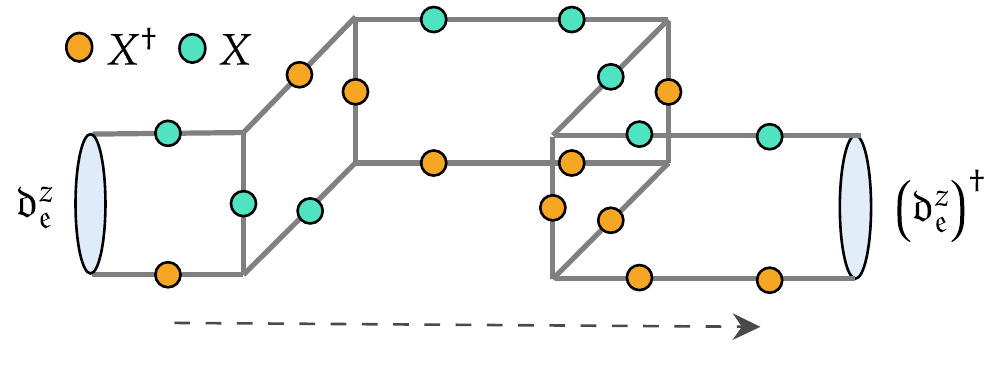}
	\caption{\label{fig:elec_dipoles} An example of a string operator which moves a $\mfd^z_\mfe$ lineon dipole (the lattice is not shown for ease of visualization). The pattern of Hermitian conjugates can be determined by comparing to the pattern appearing in the definition of the $B_c$ operator. }
\end{figure}

 We will denote a ``minimal-strength'' lineon dipole by $\mfd^a_\mfe$, where $\uva$ is the normal vector to the plane of motion. Each $\mfd^a_\mfe$ is built of an $\mfe^b$ and an $(\mfe^b)^\da$ separated by one lattice spacing in the $\uva$ direction, where $b\neq a$. There are no local processes which turn a $\mfd^a_\mfe$ dipole into a $\mfd^{b\neq a}_\mfe$ dipole. 
The string operators which move lineon dipoles in their planes of motion are ``wireframe operators'', and are given by products of $X_\lij$ operators along the edges of rectuangular prism-like shapes. The best way to understand the geometry of these string operators is with a picture -- see figure \ref{fig:elec_dipoles} for an example. {In terms of planar $\zn$ charges as described above, $\mfd^a_\mfe$ carries $\zn$ charges $+1$ and $-1$ in the two planes normal to $\uva$ in which the constituent lineon excitations reside.}

\begin{figure}
	\includegraphics{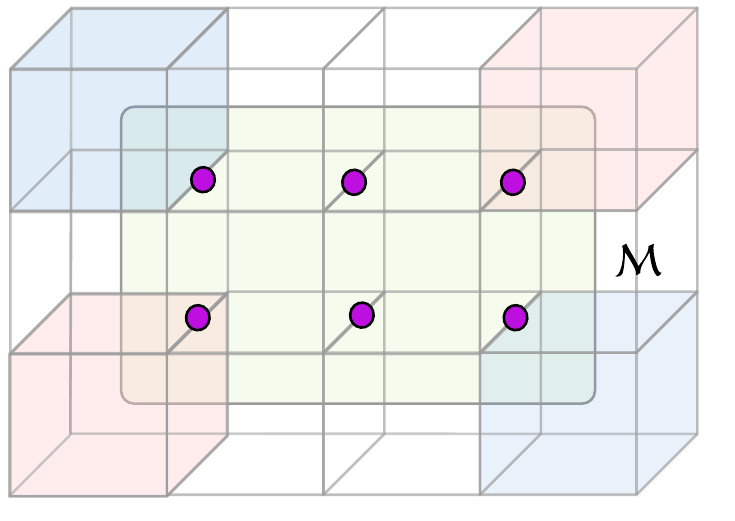} 
	\caption{\label{fig:fracton_membrane} A membrane operator supported on the green membrane $\mcm$, which is formed as a product of $Z_\lij$ operators on the links marked by the purple circles. Acting on a ground state, this operator creates four fractons living on the shaded cubes. Blue cubes have charge $+1$, and red cubes have charge $-1$ (mod $N$). } 
\end{figure}

The second sector of excitations consists of fractons and their composites. Isolated fractons arise as violations of a single $B_c$ term in $H_{XC}$, and are created with $Z_\lij$ operators. Acting with $Z_\lij$ on a ground state creates a configuration of four fractons at the cubes that contain $\lij$ as an edge, with the $\zn$ charges of the fractons arranged in a ``quadrupolar'' pattern.
More generally, acting with a product of $Z_\lij$ operators over all links $\lij$ dual to a rectangular membrane $\mcm$ of dual-lattice plaquettes creates one fracton at each of the four corners of the membrane $\mcm$, as shown in figure \ref{fig:fracton_membrane}.  
A membrane operator supported on $\mcm$ obeys $\zn$ commutation relations with lineon string operators that intersect $\mcm$; this fact enables certain types of statistical braiding-like processes to be defined between fractons and lineons.\cite{pai2019fracton}

{A single isolated fracton cannot move without creating additional excitations. In more detail, and more generally, the mobility of any collection of fractons can be understood by thinking of each fracton as carrying a conserved $\zn$ charge in the three planes normal to $\uvx$, $\uvy$ and $\uvz$ that intersect at its position.\cite{vijay2016fracton, pai2019fracton} As for lineons, local operators carry vanishing planar $\zn$ charges.  It is important to note that the planar charges carried by lineons and fractons are distinct; these are two independent sectors of excitations.}

\begin{figure}
	\includegraphics[width=.48\textwidth]{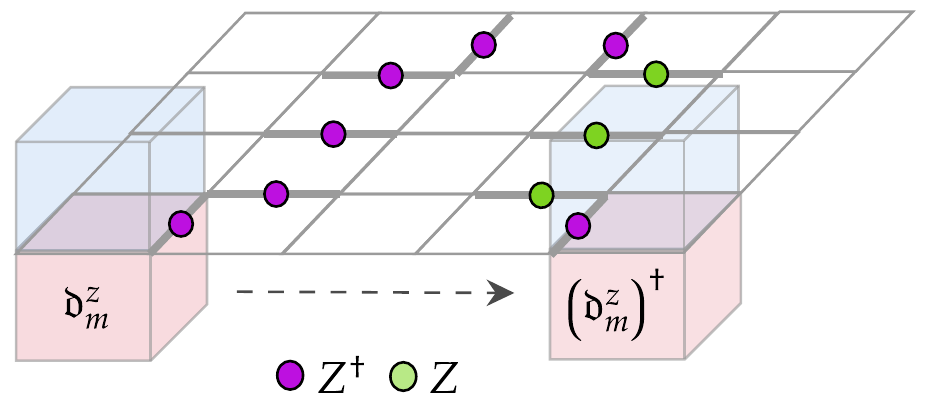}
	\caption{\label{fig:mag_dipoles} An example of a string operator which moves a $\mfd^z_\mfe$ fracton dipole from one location to another. } 
\end{figure}

 Similarly to the lineon sector, dipolar\footnote{As with lineons, we will continue to abuse terminology by referring to these bound states as ``dipoles''.} bound states of fractons can freely move in the plane normal to their dipole moment. More specifically, by a fracton dipole we mean a pair of fracton and anti-fracton excitations separated along the $x$-, $y$- or $z$-axis. The separation between the two constituent excitations cannot change by a local process, and moreover the dipole cannot move along the direction of its moment.  We will denote a ``minimal-strength'' fracton dipole capable of moving in a plane normal to $\uva$ as $\mfd^a_\mfm$. The constituent fracton and anti-fracton excitations in $\mfd^a_\mfm$ are separated by a single lattice spacing along the $\uva$ direction. There is no local process which can convert a $\mfd^a_\mfm$ dipole to a $\mfd^{b\neq a}_\mfm$ dipole. Fracton dipoles are moved using products of $Z_\lij$ operators along links lying within the plane of their motion, with an example shown in figure \ref{fig:mag_dipoles}. {Moreover, just as for lineon dipoles, $\mfd^a_\mfm$ carries $+1$ and $-1$ planar $\zn$ charges in the planes normal to $\uva$ that contain the constituent fracton and anti-fracton excitations. As a consequence, the mobility restrictions on fracton and lineon dipoles -- if no other types of excitations are present -- are \emph{exactly} identical.}

Due to the relation between the XC model and coupled layers of $\zn$ gauge theories,\cite{ma2017fracton,vijay2017isotropic} we will refer to lineon excitations and their composites as ``electric'' excitations, and fracton excitations and their composites as ``magnetic'' excitations.

\section{Fracton dipole condensation \label{sec:frac_dip}} 

We now turn to discussing the condensation of fracton dipoles in the $\zn$ XC model. We will first identify the phases that result upon condensing the fracton dipoles, using a rather explicit generalized gauge theory construction with some similarities to the membrane-net model of Ref. \onlinecite{slagle2019foliated}. Later in section \ref{sec:2dphase_transitions_fracton} we will examine the critical points that occur at the condensation transition.

\ss{Condensed phases and dual gauge theories \label{sec:frac_dip_phases}}

Before getting into details, let us ask on general grounds what types of phases we expect to obtain upon condensing various types of fracton dipoles, while leaving single fractons uncondensed.

\begin{figure}
	\includegraphics[width=.45\textwidth]{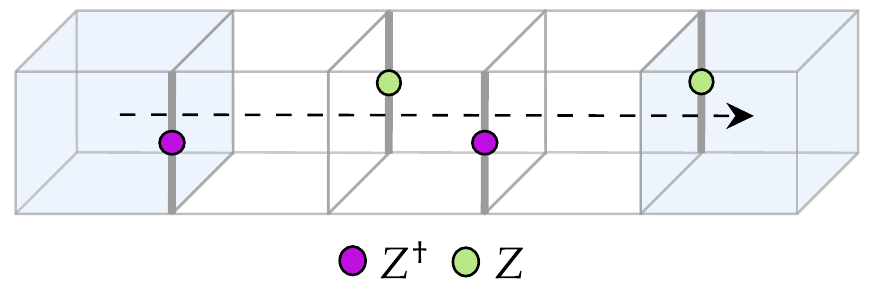}
	\caption{\label{fig:dipole_condensed_string_operator} An example of a string operator which moves an isolated fracton along the $\uvx$ direction in the presence of a condensate of $\mfd^x_\mfm$ dipoles. 
}
\end{figure}

Consider first condensing all three types of fracton dipoles. Since fracton dipoles have nontrivial statistical phases with lineons which move in their plane of motion, all of the lineons will be confined in the condensed phase. Nevertheless, the resulting phase is not trivial. Indeed, single isolated fractons remain as well-defined deconfined excitations, as there remains no local process which can create an isolated fracton. Furthermore by absorbing dipoles from the condensate, isolated fractons may now move freely in all three dimensions, via processes like the one shown in figure \ref{fig:dipole_condensed_string_operator}. It is therefore natural to identify the resulting phase with deconfined 3d $\zn$ gauge theory.\footnote{In fact, in this case standard 3d $\zn$ gauge theory is the only possibility---the set of 3d topological orders with bosonic gauge charges which have fusion rules given by a group $G$ are fully enumerated by 3d Dijkgraaf-Witten $G$-gauge theories (aka twisted $G$-gauge theories),\cite{lan2018classification} which in the case of $G = \zn$ are classified by $H^4(\zn;U(1)) = H^5(\zn;\zz) = \zz_1$, meaning that $\zn$ gauge theory is the only possibility. }

We can also contemplate condensing only a certain subset of the $\mfd_\mfm^a$ dipoles.
First, consider condensing just $\mfd_\mfm^z$. The condensate allows isolated fractons to move in the $\uvz$ direction, so that they become lineons. The condensate confines $\mfe^x$ and $\mfe^y$ lineons, while $\mfe^z$ lineons remain deconfined. As before, the $\mfe^z$ lineons can pair into $\mfd^{x,y}_\mfe$ dipoles, which are planons.

 Next, consider condensing both $\mfd^x_\mfm$ and $\mfd^y_\mfm$. This condensate leaves $\mfd^z_\mfe$ dipoles deconfined, but confines the rest of the lineon sector. In the fracton sector, isolated fractons can move in both the $\uvx$ and $\uvy$ directions, but remain immobile in the $\uvz$ direction. The deconfined excitations are thus single fractons and $\mfd^z_\mfe$ dipoles, both of which are capable of moving only within planes normal to $\uvz$. Since both of these excitations are bosons, the \ex content of the phase obtained this way is the same as a stack of 2d deconfined $\zn$ gauge theories, with the fractons corresponding to the $e$ \exs and the $\mfd^z_\mfe$ dipoles corresponding to the $m$ excitations. We will see later that this is indeed the correct identification of the condensed phase.

\begin{figure} \centering
	\includegraphics{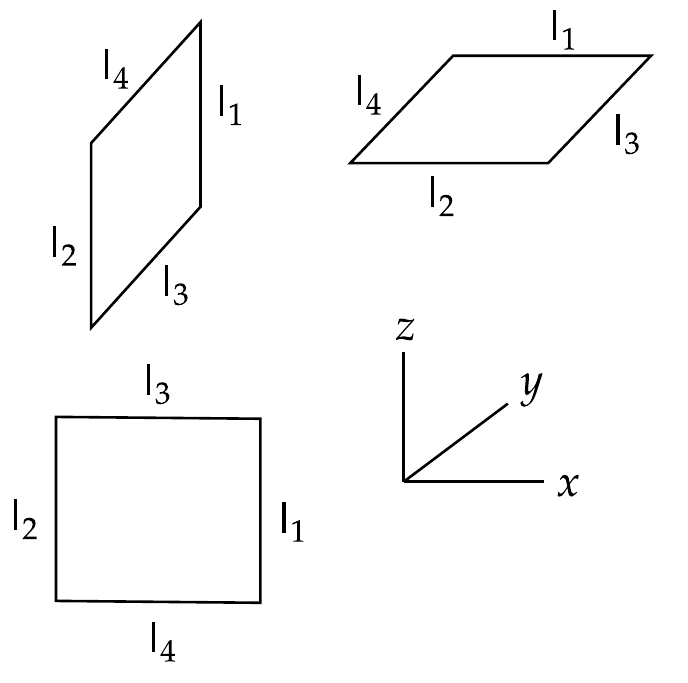}
	\caption{\label{fig:std_labeling} Conventions for indexing the links on plaquettes. All plaquettes with a given unit normal are labeled identically.} 
\end{figure}

Now we will show how these expectations can be demonstrated explicitly. For simplicity of notation, in what follows we will specialize to the $N=2$ case. The case of general $N$ is similar, and is presented in appendix \ref{sec:frac_dip_zn}. We will first focus on the case where all species of fracton dipoles condense.

Since in the condensed phase fractons become deconfined mobile particles, it is helpful to reformulate the XC model as a generalized gauge theory, where fractons are the gauge charges. To do this it is convenient to perform a duality mapping on the XC model, with the dual representation being formulated in terms of spins living on the dual lattice. In what follows, we will use {sans serif symbols to denote elements of the dual lattice}, with $\sfi,\sfl,\sfp,$ and $\sfc$ standing for dual vertices, links, plaquettes, and cubes, respectively. 
Our duality mapping involves matter qubits $\sfx_\sfi,\sfz_\sfi$ living on the vertices of the dual lattice, and gauge qubits $\mcx_\sfp,\mcz_\sfp$ living on the plaquettes of the dual lattice. Sometimes it will be convenient to use $\sfl$ to refer to a unit vector along the given link, and $\sfp$ to refer to a unit vector normal to the given plaquette, so that for instance $\sfl \prl \uvz$ denotes a link parallel to $\uvz$, while $\sfp \prl \uvz$ denotes a plaquette with normal in the $\uvz$ direction. 

The duality works in much the same way as that between the $\zt$ toric code and $\zt$ gauge theory coupled to Ising matter,\cite{kitaev2003fault} with the operators mapping as 
\be X_l \ra \mcx_\sfp,\qquad Z_l \ra \mcz_\sfp \prod_{\sfi \in  \sfp} \sfz_\sfi,\ee 
where $\sfp$ is the dual lattice plaquette dual to $l$, and the product is over the four corners of $\sfp$. In the dual formulation, there is a Gauss' law constraint at each vertex $\sfi$
\be \sfx_\sfi = \prod_{\sfp \ni  \sfi } \mcx_\sfp,\ee 
where the product is over the twelve dual plaquettes meeting $\sfi$ at a corner. The Hamiltonian in this representation is 
\be H_{XC}' = - K \sum_\sfi \sfx_\sfi - g \sum_{\sfc,a} A^a_\sfc.\ee 
Here $A^a_\sfc$ is a product of $\mcz_\sfp$ over four of the six plaquettes in the cube $\sfc$, excluding the two plaquettes normal to the $\uva$ direction. 

We will now modify this model to include a separate field for fracton dipoles. As such dipoles naturally live on the links of the dual lattice, we introduce a qubit on each link with Pauli operators $X_\sfl,Z_\sfl$ (not to be confused with the Pauli operators in the original formulation of the XC model, which live on the direct lattice), where $X_\sfl$ is the dipole number operator and $Z_\sfl$ creates a fracton dipole at the link $\sfl$. 

Since single fractons carry gauge charge, a dipole at $\sfl$ should carry gauge charge on the vertices at the two ends of $\sfl$. We therefore modify the Gauss law constraint to be 
\be \label{z2_full_gauss} \sfx_\sfi  \prod_{\sfl \ni i} X_{\sfl} = \prod_{\sfp \ni \sfi} \mcx_\sfp,\ee 
where the products are over all the links and plaquettes neighboring the vertex $\sfi$. 
With this Gauss law two fractons at adjacent vertices can be converted into a dipole via the gauge-invariant operator $\sfz_\sfi Z_\lijd \sfz_\sfj$, and two dipoles with parallel dipole moments can be created on opposite edges $\sfl,\sfl'$ of a plaquette $\sfp$ via the gauge-invariant operator $Z_\sfl \mcz_\sfp Z_{\sfl'}$. 

To analyze what happens when we condense the dipoles, it will be notationally helpful to fix a standard set of links $\sfl_1,\dots ,\sfl_4$ for each plaquette of a given orientation. Our conventions for each plaquette are shown in \ref{fig:std_labeling}, and are such that for any plaquette $\sfl_1$ is always parallel to $\sfl_2$, and $\sfl_3$ is always parallel to $\sfl_4$. 

We now consider the following Hamiltonian:
\bea \label{fracdip_ham} H_{con} & = - K \sum_\sfi \sfx_\sfi - g \sum_{\sfi,a} A^a_\sfi \\ & - h\sum_\sfl X_\sfl - \l \sum_{\sfp} (Z_{\sfl_1} \mcz_\sfp Z_{\sfl_2} + Z_{\sfl_3} \mcz_\sfp Z_{\sfl_4} ).\eea 
The last term proportional to $\l$ is a kinetic energy term for the dipoles, and allows dipoles on links parallel to $\uva$ to move in planes normal to $\uva$. 

When $h/\l \gg1$, we are in the regime where both the dipoles and fractons are gapped, and the gauge field $\mcz$ is deconfined. In this limit we are clearly in the same phase as $H_{XC}$. In the following we will instead be interested in the limit $h/\l \ll 1$, as in this limit the dipoles are condensed. In this limit the second term in \eqref{fracdip_ham} can be dropped without changing the low-energy physics. This is because $A^a_\sfi$ can be obtained by taking products of the dipole kinetic terms, so that if these terms are set to $1$ then so are the $A^a_\sfi$ terms. 

We will now show that in the phase where dipoles are condensed, the Hamiltonian \eqref{fracdip_ham} can be transformed into the standard Hamiltonian of $\zt$ gauge theory, plus an unimportant decoupled paramagnet arising from the plaquette gauge fields. {For simplicity and to work at a solvable point within the dipole-condensed phase, we set $h = g = 0$.}	Note that if we ignore the plaquette variables $\mcx_\sfp$, the Gauss law \eqref{z2_full_gauss} becomes the ordinary $\zt$ gauge theory Gauss law constraint. Our strategy will therefore be to 
arrive at the $\zt$ gauge theory by performing a unitary transformation designed to eliminate the $\mcx_\sfp$ variables appearing in \eqref{z2_full_gauss}.

To begin, for each plaquette, define the CNOT-like operators
\be U_{\sfp} \equiv \( \frac{1 + Z_{\sfl_1}Z_{\sfl_2}}{2}\) + \( \frac{1- Z_{\sfl_1} Z_{\sfl_2}}2\) X_\sfp.\ee 
Note that $U_{\sfp} = U^\da_{\sfp}$ is unitary, and is {\it not} gauge invariant --- thus we can use it to transform both the Hamiltonian and the gauge constraint. 

The unitary $U_{\sfp}$ conjugates the various Pauli operators as 
\bea \label{uijids} U_{\sfp} \mcz_\sfp U_{\sfp} & = \mcz_\sfp Z_{\sfl_1}Z_{\sfl_2} \\ 
U_{\sfp} X_{\sfl_{1,2}} U_{\sfp} & = X_{\sfl_{1,2}} \mcx_\sfp \\ 
U_{\sfp} X_{\sfl_{3,4}} U_{\sfp} & = X_{\sfl_{3,4}} \text{,} \eea 
with other conjugations trivial.

{Noting that $[U_{\sfp}, U_{\sfp'}] = 0$,} we now perform a unitary conjugation by the operator 
\be \mcu \equiv \prod_\sfp U_{\sfp}.\ee 
{While the support of $U_{\sfp}$ operators on adjacent plaquettes does overlap, $\mcu$ can be written as a depth-two quantum circuit by dividing the plaquettes into two subsets.}
The transformed Hamiltonian  then takes the form
\bea \label{tform_hp} H'_{con}& = \mcu  H_{con}\mcu \\ 
& = -K \sum_\sfi \sfx_\sfi -\l\sum_\sfp \mcz_\sfp - \l \sum_\sfp \mcz_\sfp \prod_{\sfl \in \sfp} Z_\sfl\eea 
Here the first term proportional to $\l$ comes from conjugating terms which hop dipoles between $\sfl_1$ and $\sfl_2$ links, while the second comes from terms which hop dipoles between $\sfl_3$ and $\sfl_4$ links. The product in the final term is over the four links in the perimeter of $\sfp$.

To compute the transformation of the gauge constraint, note that for a given vertex $\sfi$, for each of the twelve plaquettes meeting $\sfi$, one of the links $\sfl_1,\sfl_2$ has $\sfi$ as an endpoint. When the $X_{\sfl}$ operator on this link is conjugated by $U$, it produces by \eqref{uijids} an $\mcx_\sfp$ operator, which cancels one of the $\mcx_\sfp$'s appearing in the Gauss law. The transformed gauge constraint is therefore 
\be \label{gaugegauss} \sfx_\sfi = \prod_{l \ni i} X_\sfl,\ee 
which is obtained by conjugating both sides of \eqref{z2_full_gauss} by $\mcu$. This is exactly the gauge constraint of $\zt$ gauge theory. 

It only remains to deform $H'_{con}$ to the standard solvable Hamiltonian for the deconfined phase of $\zt$ gauge theory. This can obviously be done without closing the gap simply by freezing out the now gauge-invariant $\mcz_\sfp$ degrees of freedom in \eqref{tform_hp}, which are gapped and have no dynamics. More precisely, we may introduce a parameter $s\in [0,1]$ and write 
\bea H'_{con}(s) & \equiv  -K\sum_\sfi \sfx_\sfi -\l\sum_\sfp  \mcz_\sfp  \\ & \qquad - \l \sum_\sfp [(1-s)\mcz_\sfp+s] \prod_{\sfl\in \sfp} Z_\sfl.\eea 
Clearly at $s=0$ we have $H'_{con}(0) = H'_{con}$, the ground state does not change as a function of $s$, and the gap remains open. At $s=1$ we have 
\be H'_{con}(1) =  -K\sum_\sfi \sfx_\sfi -\l\sum_\sfp \mcz_\sfp - \l \sum_\sfp  \prod_{\sfl \in \sfp} Z_\sfl.\ee
This is nothing but the usual $\zt$ gauge theory in its deconfined phase, stacked with a trivial paramagnet. 

Note that the gauge field variables in the condensed phase are precisely the dipole creation operators. This can be argued intuitively: in the condensed phase, single fractons can only move with the help of string operators built from dipoles, and in the gauge theory description these strings become the Wilson lines which render isolated fractons gauge invariant. 

\ss{Condensing a subset of fracton dipoles}

Identifying the nature of the phases obtained when condensing only a subset of the fracton dipoles can be done similarly. We give the details here in the case $N = 2$.

\sss{Condensing a single orientation of dipoles}
 Suppose first that we condense dipoles with only a single orientation of dipole moment, which given our conventions is most conveniently chosen to be $\uvz$. This can be done simply by using the Hamiltonian \eqref{fracdip_ham}, but only including $X_{\sfl}$, $Z_{\sfl}$ dipole fields, and the corresponding hopping terms, for links parallel to $\uvz$. We thus consider the Hamiltonian 
\bea H_{con;z} & = -K\sum_\sfi \sfx_\sfi - g \sum_\sfc (A^x_\sfc + A^y_\sfc) \\ \qquad & -  \l \sum_{\sfp \perp \uvz } Z_{\sfl_1} \mcz_{\sfp} Z_{\sfl_2}  \text{,}
\eea 
where we have left out $A^z_\sfc$ in the second term on the grounds that it can be written as a product of dipole kinetic energy terms, and where the last term involves a sum over plaquettes with normal vectors perpendicular to $\uvz$. Note that the pairs $\sfl_1,\sfl_2$ are parallel to $\uvz$ for both orientations of plaquettes with $\sfp \perp \uvz$; see figure \ref{fig:std_labeling}.  
The Gauss' law constraint is an appropriate anisotropic version of \eqref{z2_full_gauss}, viz. 
\be \label{z2_anisotropic_gauss} \sfx_\sfi \prod_{\sfl \prl \uvz \ni \sfi} X_\sfl = \prod_{\sfp \ni \sfi}  \mcx_\sfp,\ee 
where the product over $\sfl$ on the left-hand side is over the two $z$-axis links $\sfl$ that touch the vertex $\sfi$.

We now perform a unitary transformation with the operator $\mcu_z = \prod_{\sfp \perp \uvz} U_\sfp$, which is the same as the operator $\mcu$ defined above, except for the omission of plaquettes with unit normal along $\uvz$. After this conjugation, we have 
\bea 
H'_{con;z} & = \mcu_z H_{con;z} \mcu_z \\ 
& =  - K \sum_\sfi \sfx_\sfi - g \sum_\sfc (A^x_\sfc + A^y_\sfc) \prod_{\sfl\prl \uvz \in \sfc} Z_\sfl -\l \sum_{\sfp \perp \uvz} \mcz_\sfp \text{.}
\eea 
Importantly, the second term is modified by the transformation, which adds a product of $Z_{\sfl}$ over the four $z$-axis links $\sfl$ contained in the cube $\sfc$.  This can clearly be deformed to
\bea \label{hcpp}
H''_{con;z}  & = -  K\sum_\sfi \sfx_\sfi  -g \sum_\sfc [\prod_{\sfp \prl \uvz \in \sfc} \mcz_\sfp] [\prod_{\sfl\prl \uvz \in \sfc} Z_\sfl] - \l \sum_{\sfp \perp \uvz} \mcz_\sfp  \text{,}
\eea 
where the first product in the second term is over the two plaquettes on the boundary of the cube $c$ with normal vectors along $\uvz$. 
Under conjugation by $\mcu_z$, the Gauss' law transforms to
\begin{equation}
\sfx_\sfi \prod_{\sfl\prl \uvz \ni \sfi} X_{\sfl} = \prod_{\sfp\prl\uvz \ni \sfi} \mcx_\sfp   \text{.}
\end{equation}
Here the product on the right-hand side is over the four plaquettes with normals along $\uvz$ that meet $\sfi$ at a corner.

The Hamiltonian \eqref{hcpp} together with the above Gauss constraint is nothing but the ``anisotropic model with lineons and planons'' discussed in Sec.~5.7 of Ref. \onlinecite{shirley2019fractional} (mapped to a generalized gauge theory in the usual way), with the two types of lineon excitations identified in Ref. \onlinecite{shirley2019fractional} corresponding here to single fractons and $\mfe^z$ lineons. 

\sss{Condensing two orientations of dipoles}

Consider now condensing dipoles with dipole moments oriented along $\uvz$ and $\uvx$. By exchanging dipoles with the condensate, the fractons in the condensed phase will be able to move along both the $\uvz$ and $\uvx$ directions, and hence the natural expectation is to identify the condensed phase with a decoupled stack of $\zt$ gauge theories. 

This is indeed what happens. We include $X_{\sfl}$, $Z_{\sfl}$ for $\sfl$ parallel to $\uvz$ and $\uvx$, but not $\uvy$. The Gauss' law is thus
\be \sfx_\sfi \prod_{\sfl \prl \uvz, \uvx \ni \sfi} X_\sfl = \prod_{\sfp \ni \sfi}  \mcx_\sfp,\ee 
where the product on the left-hand side is over the $x$ and $z$ links touching $\sfi$.
For the Hamiltonian we take \eqref{fracdip_ham} and include only kinetic terms for dipoles oriented along the $\uvz,\uvx$ directions. We have, referring to figure \ref{fig:std_labeling},
\bea H_{con;zx} & = -K\sum_\sfi \sfx_\sfi - \l \sum_{\sfp \prl \uvx, \uvz} Z_{\sfl_1} \mcz_\sfp Z_{\sfl_2} \\ 
& \qquad - \l \sum_{\sfp \prl \uvy} \mcz_\sfp ( Z_{\sfl_1}Z_{\sfl_2}+Z_{\sfl_3}Z_{\sfl_4}).\eea
We have again set $h = g = 0$, noting that the $A^a_{\sfc}$ terms can be obtained as products of the dipole kinetic energy terms that are present. Conjugating with $\mcu$ as in the analysis of the case where all dipoles condense, we obtain 
\be \mcu H_{con;zx}\mcu  =  -K\sum_\sfi \sfx_\sfi - \l \sum_{\sfp} \mcz_\sfp - \l \sum_{\sfp \prl \uvy} \mcz_\sfp \prod_{\sfl\in \sfp} Z_\sfl,\ee 
while the Gauss law becomes
\begin{equation}
\sfx_{\sfi} = \prod_{\sfl \prl \uvz, \uvx \ni \sfi} X_{\sfl} \text{.}
\end{equation}
After eliminating the now-trivial $\mcz_\sfp$ degrees of freedom, this indeed describes a set of decoupled deconfined $\zt$ gauge theories, stacked along the $\uvy$ direction. %

\ss{Phase transitions \label{sec:2dphase_transitions_fracton}}

We now turn to studying the critical points that occur between the $\zn$ X-cube phase and the condensed phases described above. We will first focus on what happens when we only condense fracton dipoles with a single orientation of dipole moment, which we take to be along $\uvz$. 

To analyze the critical points we can ignore the fracton matter spins $\sfx,\sfz$, as single fractons remain gapped across the transition. In addition, excitations of the generalized gauge field are gapped across the transition, so we may also ignore fluctuations of the generalized gauge field ${\mathcal Z}_{\sfp}$. Indeed, we may set ${\mathcal Z}_{\sfp} = 1$ and focus on an effective Hamiltonian for the dipole matter alone, provided that we remember to only consider correlation functions of gauge-invariant operators. Since for now we are only interested in condensing $\mfd^z_\mfm$ dipoles, we only need to retain the corresponding kinetic terms. Thus we can consider the Hamiltonian 
\be \label{hcp} H_{stack} = -h \sum_{\sfl\prl\uvz} X_{\sfl} - \l \sum_{\sfp \perp \uvz} Z_{\sfl_1}^\da Z_{\sfl_2} + h.c,\ee 
where the $\sfl_1,\sfl_2$ links are parallel to $\uvz$ (see figure \ref{fig:std_labeling}), and where we are again working with $\zn$ spins. Since only the spins on the links parallel to $\uvz$ enter into the above Hamiltonian, we see that $H_{stack}$ is simply a decoupled stack of 2d $\zn$ clock models, with one clock model at each $z$-coordinate of the lattice. %

In this effective matter theory, {the $\zn$ planar conservation laws governing mobility of lineons (see Sec.~\ref{sec:preliminaries}) lead to a subsystem symmetry.} In particular, the number of dipoles in each $xy$
plane is separately conserved modulo $N$.\footnote{Strictly speaking, this conservation only holds for local terms; non-local processes that add a dipole to every plane are allowed, because each dipole carries two opposite-sign $\zn$ charges in neighboring $xy$-planes. A more general statement is the the difference in number of dipoles between any two planes is conserved.} From a formal perspective, this symmetry arises from the fact that any local term in the matter theory can be consistently coupled to the generalized gauge field if and only if it is symmetry-invariant. %

The critical point of the Hamiltonian \eqref{hcp} is then the same as a decoupled stack of critical 2d $\zn$ clock model layers, with the added restriction 
-- coming from gauge invariance -- of the effective subsystem symmetry. The allowed operators are precisely those that are invariant under independent $\zn$ transformations on each layer.

\sss{$N=2$}

What happens at the critical point depends on the value of $N$. 
First, consider the case of $N=2$. Here the decoupled critical point is built from Ising$^*$ CFTs on each layer, where as usual the $*$ denotes the restriction to the $\zt$-neutral part of the spectrum. The most relevant way to couple different layers together is through their energy operators, via a perturbation to the fixed-point action of the form 
\be \d S = \sum_{\a\neq\b} g_{\a,\b}\int d^2x \, d\tau \, \ep_\a \ep_\b,\ee 
where $\ep_\a$ is the energy operator on layer $\a$. Since the scaling dimension of the energy operator in the 2d Ising model is $\De_\ep \approx 1.41 < 3/2$,\cite{poland2019conformal} these couplings are slightly relevant at the decoupled fixed point. In appendix \ref{layeredIsing} we argue that there is in fact no stable fixed point to this RG flow, with the result being an instability towards a  first-order transition. 

\sss{$N=3$}
When $N=3$, we have a 3-state Potts model on each layer. The transition in the 3-state Potts model {in three space-time dimensions} is well-known to be first order, and hence we do not obtain any stable fixed points in this case. 

\sss{$N=4$}
The case of $N=4$ can be understood by noting that the $\zz_4$ clock model can be mapped to a pair of two decoupled Ising models, written in terms of two $\zt$ spins $Z_1,Z_2$ (see e.g. \onlinecite{ortiz2012dualities}).\footnote{
To see this, we write the $\zfo$ clock matrix $Z$ in terms of two $\zt$ Pauli matrices $Z_1\equiv \s^z \tp \unit,Z_2 \equiv \unit \tp \s^z$ as
\be Z = \frac1{\sqrt2} (e^{i\pi /4} Z_1 + e^{-i\pi/4} Z_2).\ee 
The real matrix $X + X^\da$ also has a simple representation in terms of $X_1 \equiv \s^x \tp \unit$ and $X_2 = \unit \tp \s^x$, with 
\be X + X^\da = X_1 + X_2.\ee 
It is then easy to check that in terms of the $\zt$ variables, the $\zz_4$ clock model Hamiltonian splits as $H_1+H_2$, where $H_{1}$ ($H_2$) is an Ising chain Hamiltonian for the $Z_1$ ($Z_2$) variables.  }
The (gauged) $\zz_4$ symmetry acts in this representation as 
\be \zz_4 \, :\, Z_1 \ra Z_2,\qquad Z_2 \ra Z_1X_1,\ee 
and as such at the critical point it exchanges the two energy operators $\ep_1$ and $\ep_2$. 
The operator $\ep_1 + \ep_2$ is therefore  gauge-invariant, and since the Ising models are decoupled $\ep_1 + \ep_2$ has the same dimension as the energy operator in a single Ising model. Therefore according to the discussion of the $N=2$ case above, the decoupled fixed point is again unstable.

\sss{$N>4$}

When $N>4$, we describe the critical point on each layer with the action 
\bea \label{rel_2d_action}S & = \int d\tau \, d^2x\, \Big( |\p\psi|^2  +t|\psi|^2 + \frac{u }{4} |\psi|^4 \\ & \qquad +  \frac{g\La^{4-N}}{N!} (\psi^N + (\psi^*)^N) + \cdots\Big),\eea
with $\psi$ a complex scalar field.
It is known from Monte Carlo simulations that the $\zn$ anisotropy term is irrelevant as long as $N>4$,\cite{scholten1993critical,hove2003criticality} so that the phase transitions on each layer are in the universality class of the 2+1D XY$^*$ model. The most relevant gauge-invariant coupling between the layers is the energy-energy coupling. The energy operator in this case is known to have scaling dimension\cite{poland2019conformal}
\be \Delta_\ep \approx 1.51 > 3/2,\ee 
and so the decoupled fixed point consisting of a decoupled stack of 2+1D XY$^*$ models is (barely) stable.

\sss{No particle-hole symmetry}

For $N > 2$, the discussion above strictly speaking pertains only to models which possess a $\zt$ particle-hole symmetry $C$, which acts on the spins of the $\zn$ clock models (as well as on the original spins of the XC model) by conjugation with the $N\times N$ matrix $C$, where 
\be C = \bpm 1 &&& \\ 
&&&1 \\ &&\scalebox{-1}[1]{$\ddots$} & \\ &1&& \epm ,\qquad C \sfz C = \sfz^\da,\quad C \sfx C = \sfx^\da.\ee 
In terms of the coarse-grained field $\psi$ employed above in \eqref{rel_2d_action}, this sends $C : \psi \ra \psi^*$.

This symmetry can be broken for $N>2$ by letting the parameters $h,\l$ in \eqref{hcp} be complex. %
We are then prompted to consider an action on each layer of the form  
\bea \label{nonrel_2d_action}S & = \int d\tau \, d^2x\, \Big(  \psi^*\p_\tau \psi +\frac{1}{2m}|\D \psi|^2  +t|\psi|^2 + \frac{u }{4} |\psi|^4 \\ & \qquad +  \frac{g\La^{4-N}}{N!} (\psi^N + (\psi^*)^N) + \cdots\Big),\eea
where the $C$-breaking first term, which leads to non-relativistic $z=2$ scaling, is included instead of $|\p_\tau\psi|^2$ on the grounds of it being more relevant. Since 2+1D is the upper critical dimension for the above model, the RG flow can be accurately computed through a perturbative analysis of the above action \eqref{nonrel_2d_action}.

The non-relativistic nature of the theory means that the 1-loop computation of the $\b$ functions is exact,\footnote{For example, in momentum-shell regularization with a cutoff $\L$, the only diagrams depending on $\ln(\L)$ are those built from concatenations of one-loop bubbles, meaning that the full beta functions are determined by the one-loop terms.} giving
\be \beta_u = -u^2/2,\qquad \beta_g = (4-N)g -ug/2.\ee
The anisotropy is therefore relevant at $N=3$, marginally irrelevant at $N=4$, and irrelevant for $N>4$. When $N=3$ Ginzburg-Landau theory predicts a first-order transition due to the shape of the potential for $\psi$, and so this case can be ignored. For $N\geq 4$ however, we obtain a nontrivial critical theory describable by the non-relativistic XY$^*$ model.

We therefore need to understand the stability of a decoupled stack of non-relativistic XY$^*$ models.\footnote{This story is very similar to the analysis of stacks of Fermi liquid $\lra$ orthogonal metal phase transitions studied in Ref. \cite{zou2016dimensional}} The most relevant gauge-invariant couplings are again those which couple the energy operators on different layers; in the present notation they read
\be \delta S = \sum_{\a,\b} g_{\a,\b} \int d\tau\,d^2x\, |\psi_\a|^2 |\psi_\b|^2.\ee 
The beta function for $g_{\a,\b}$ is similarly exactly computable from the 1-loop term, which gives 
\be \beta_{g_{\a,\b}} = -\frac12 (g_{\a,\b})^2.\ee  
Due to the non-relativistic nature of the theory, none of the $g_{\a,\b}$s mix with each other. As a result, the decoupled fixed point is stable, provided that all of the $g_{\a,\b}$s are positive.\footnote{Note that getting a stable fixed point does not require fine-tuning: we just need to that $g_{\a,\b}>0$ for all $\a,\b$; we do not require that each of the $g_{\a,\b}$ be tuned to any one particular value.}

Before moving on, we add a brief aside about a subtlety in the above discussion of stability. In the argument above, we have restricted our attention only to the $g_{\a,\b}$ couplings, which are marginal. In principle however, the irrelevant operators we have neglected may drive some of the $g_{\a,\b}$ negative during the initial stages of the RG flow. If this occurs generically, the stability of the proposed fixed point would be called into question. 

This worry is not a serious issue in the present context, however. First, note that due to the non-relativistic scaling, the only irrelevant operators which can renormalize $g_{\a,\b}$ are those containing $\psi^*_\a\psi_\a\psi^*_\b\psi_\b$ and no other field operators, the least irrelevant of which is $(\psi_\alpha^*\nabla_i \psi_\alpha) (\psi_\beta^* \nabla^i \psi_\beta)$, which has an RG eigenvalue of $-2$. Unless the bare value of this term is significantly larger than that of $g_{\alpha,\beta}$, $g_{\a,\b}$ will never be driven negative during the flow. 
Therefore if we make the (reasonable) assumption that the bare values of these irrelevant operators are not significantly larger than the bare values of the appropriate $g_{\a,\b}$, the above fixed point is indeed stable.\footnote{It is instructive to compare the story here with the Kohn-Luttinger instability that occurs in Fermi liquids.\cite{shankar1994renormalization} In that scenario, irrelevant operators lead to instabilities by renormalizing certain marginal couplings. This occurs however only due to the fact that the bare values of certain marginal terms are naturally exponentially smaller than the bare values of the irrelevant operators which renormalize them. In the present setting, by contrast, {the bare values of the irrelevant terms and of the marginal terms they renormalize both scale similarly with the distance between the $\a$ and $\b$ layers}.
} 

\sss{Condensing multiple species of dipoles}

So far we have only been concerned with condensing a single species of dipole, but the same analysis can be applied when simultaneously condensing multiple species. The starting point for the critical theory is simply multiple decoupled stacks of critical $\zn$ clock models, with one stack oriented along the dipole moment vector of each of the condensing dipoles. The effective matter theory again has a subsystem symmetry, where (for local terms) the number of $z$-dipoles is conserved modulo $N$ in each $xy$-plane, and similarly for $x$- and $y$-dipoles. {Dipoles with different orientations of dipole moment are separately conserved, because they carry $\zn$ planar charges in distinct planes.} The discussion of stability of the decoupled critical point is not modified, because the couplings between perpendicular planes are always less relevant than the couplings between parallel planes.

\section{Lineon dipole condensation \label{sec:2d_lineons}}

In this section we discuss what happens when lineon dipoles are condensed. 
The analysis is similar in many respects to the case of fracton dipoles treated in the previous section, although the phenomenology is slightly richer due to fact that there are three different types of lineons, as opposed to only a single type of fracton. 
As in the previous section, we will proceed by first using a generalized gauge theory to identify the nature of the condensed phases, and then later turn to discussing the nature of the phase transitions. 

\ss{Condensed phases and dual gauge theories \label{sec:phases_lineon_dipoles}}

\begin{figure}
	\includegraphics[width=.4\textwidth]{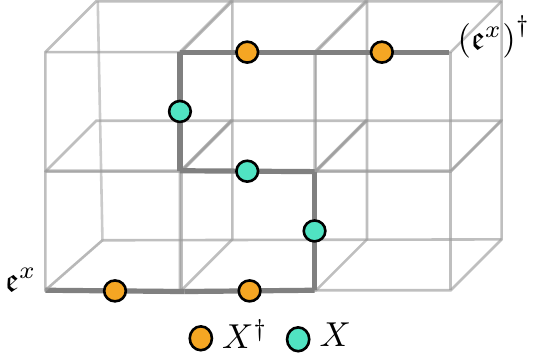}
	\caption{\label{fig:elec_dipole_condensed_string_operator} An illustration of a string operator moving an $\mfe^x$ excitation in the $\uvz$ direction in the presence of a $\mfd^z_\mfe$ condensate. The extra \exs created by the middle part of the string may all be absorbed into the condensate. }
\end{figure}

Before discussing concrete Hamiltonians, let us assess our expectations for the condensed phases. 
First consider condensing a single species of dipole, say $\mfd^z_\mfe$. In the electric sector, $\mfe^{x}$ and $\mfe^y$ lineons now become capable of moving along the $\uvz$ direction, as they can do so by absorbing dipoles from the condensate. An example of a string operator implementing this type of process is shown in figure \ref{fig:elec_dipole_condensed_string_operator}, where an $\mfe^x$ lineon moves in the $\uvz$ direction by absorbing $\mfd^z_\mfe$ dipoles from the condensate. The $\mfe^x$ and $\mfe^y$ lineons still remain distinct \exs however, as there continues to be no local process which converts an isolated $\mfe^x$ into an isolated $\mfe^y$. The electric sector thus consists of two types of $\zn$ charges, each of which are free to move in planes normal to the $\uvx$ and $\uvy$ directions.

\begin{figure}
    \includegraphics[width=.5\textwidth]{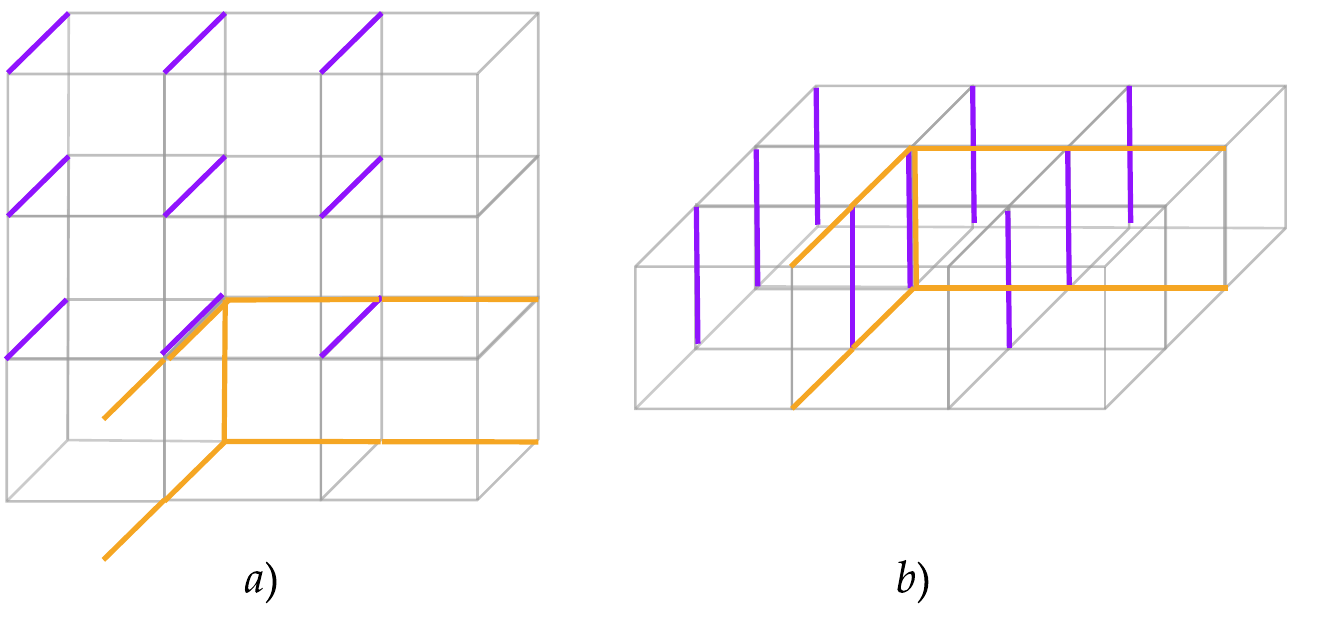}
    \caption{Processes which give tension to fracton membranes in the presence of a $\mfd^z_\mfe$ condensate. In both panels, purple links denote the support of fracton membrane operators, while orange links denote the support of condensed lineon dipole strings. a) a process giving line tension to a membrane in an $xz$ plane. b) a corner-turning process giving surface tension to a membrane in an $xy$ plane.}
    \label{fig:dipoles_and_sheets}
\end{figure}

In the magnetic sector, the condensate gives a tension to the membrane operators creating fractons at their corners in a rather subtle way. For such membrane operators lying in a $xz$-plane, the string operators of the condensed $\mfd^z_\mfe$ dipoles only fail to commute with the membrane near its edges running along the $x$-direction; an example of such a processes is shown in the left panel of Fig. \ref{fig:dipoles_and_sheets}. This means that the $x$-edges of the membrane acquire a line tension. Similarly, the $y$-edges of membranes lying in a $yz$-plane also get a line tension from the condensate. On the other hand, membranes lying in $xy$-planes get a surface tension, because the string operators of condensate dipoles lying in the same $xy$-plane anticommute with the membrane operator, if the dipole turns a corner anywhere within the area of the membrane; an example is shown in the right panel of Fig. \ref{fig:dipoles_and_sheets}. These results imply that isolated fractons, as well as $\mfd^z_\mfm$ dipoles are confined. However the $\mfd^{x,y}_\mfm$ dipoles remain deconfined.

The deconfined excitations of the theory are thus the same as that of two stacks of deconfined 2d $\zn$ gauge theories, with one stack oriented along $\uvy$ and another along $\uvx$. We will see later that the condensed theory is indeed given by two interpenetrating stacks of 2d $\zn$ gauge theories. 

Now consider condensing a second species of dipole, say $\mfd^y_\mfe$. Since $\mfd^y_\mfe $ is a bound state of the gauge charges in two adjacent layers in the gauge theory stack oriented along $\uvy$, condensing $\mfd^y_\mfe$ will turn this stack into a single 3d gauge theory, while leaving the stack oriented along $\uvx$ unchanged. We therefore obtain a 3d $\zn$ gauge theory (with gauge charge $\mfe^x$) and a stack of 2d $\zn$ gauge theories oriented along the $\uvx$ direction (with gauge charges given by $\mfe^y$ on each layer).
 
Finally, consider the phase obtained when all the $\mfd^a_\mfe$s are condensed. Each $\mfe^a$ is now free to move in all three directions, and since there is still no local process turning an $\mfe^x$ into an $\mfe^y$, a natural assumption is that the condensed phase is a deconfined 3d $\zn^2$ gauge theory.

The above logic gives plausible identifications for the condensed phases in each case, but is not completely rigorous. For example, while the electric sector in the phase where all $\mfd^a_\mfe$ condense is the same as that of $\zn^2$ gauge theory, it is not obvious how the magnetic degrees of freedom in the XC model end up organizing themselves into a $\zn^2$'s worth of loop excitations. 
 
To conclusively demonstrate that these expectations are borne out, we will employ a generalized gauge theory construction similar to that used when discussing the condensation of fracton dipoles.

 We will first focus on the case where all orientations of dipoles condense. We will furthermore specialize to the $\zt$ case for simplicity of notation; the generalization to $\zn$ is done using the same methods as employed in Appendix~\ref{sec:frac_dip_zn}. According to the discussion above, we expect that the resulting phase will be a deconfined $\zt^2$ gauge theory. To show this, it is helpful to realize the XC model as a particular limit of a Hamiltonian defined in a larger Hilbert space. Since we are aiming for a $\zt^2$ gauge theory, we will want to have two degrees of freedom on each link. In fact we will find it more convenient to work with a model containing three qubits on each link, together with a single constraint reducing the total number of degrees of freedom on each link to two.

\begin{figure} \centering
	\includegraphics{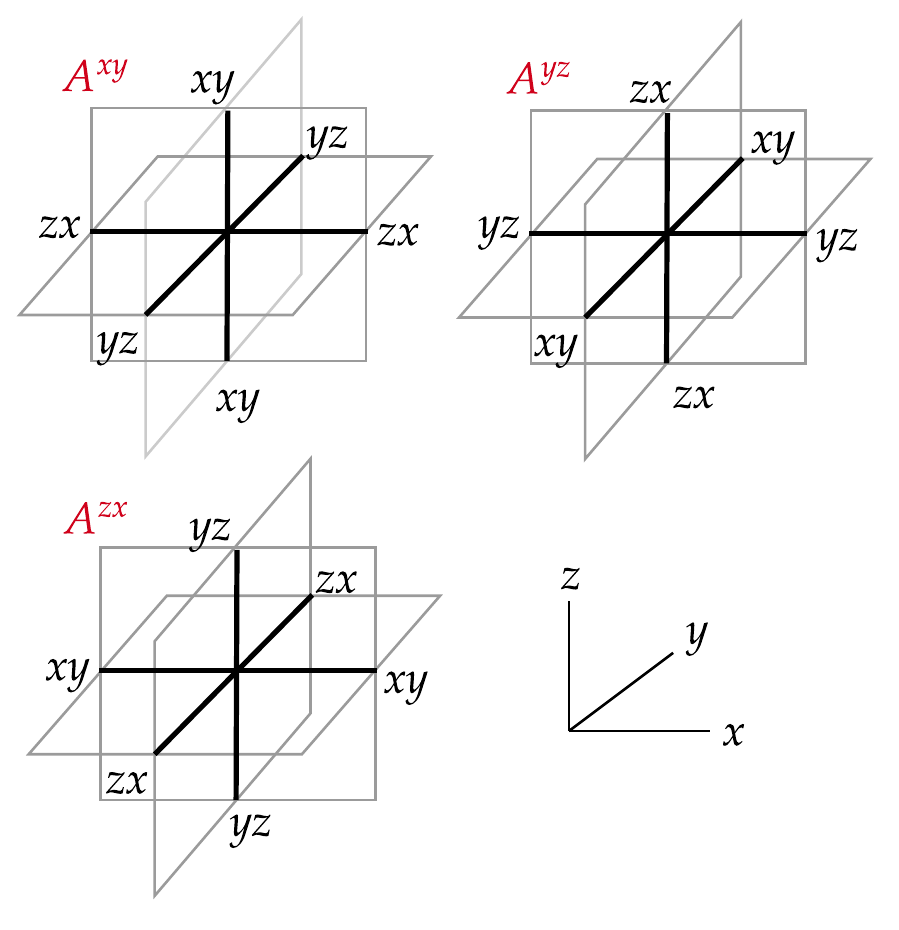}
	\caption{\label{fig:gauss_flowers} Representations of the gauge field terms appearing in the generalized Gauss laws. The operators $A^{ab}$ are defined as products of pairs of $X^cX^d$ operators over each of the black links, with $c,d$ indicated by the pairs of letters. 
	} 
\end{figure}

We thus start with a Hamiltonian defined on a Hilbert space with three qubits on each link of the direct lattice, with Pauli operators $X_l^a,Z_l^a$ ($a=x,y,z$), which are subject to the constraint 
\be \label{threez} Z^x_l Z^y_l Z^z_l = 1 \ee 
for each link $l$. 	
Within this Hilbert space, we can write a Hamiltonian that reduces to $H_{XC}$ as 
\be \label{hxcp_lineon} H_{XC}' = - g\sum_i \sum_{a\neq b} A_i^{ab} - K \sum_c \prod_{l\in c} Z^l_l
- h \sum_l X^a_l X^b_l ,\ee 
where $Z^l_l$ means $Z^a_l$ with $\uva \prl l$, and where in the last term $a \neq b$ with $\uva , \uvb \perp l$. The operators $A^{ab}_i$, $a\neq b$ are defined as 
\bea \label{eqn:Aab} A^{ab}_i = \prod_{l \perp \uva \ni i} X^b_l \prod_{l' \perp \uvb \ni i} X^a_{l'} \prod_{l''\perp \uvc \ni i} X^c_{l''},\eea 
where $c$ is neither $a$ nor $b$, and where for instance the first product is over the four links touching $i$ that are perpendicular to $\uva$. Note that $A^{a b}_i = A^{b a}_i$. An illustration of the three different types of $A^{ab}$ operators is given in figure \ref{fig:gauss_flowers}. $H_{XC}'$ preserves the constraint \eqref{threez}, since each term contains an even number of $X^a_l$ operators on each link. Moreover, all the terms in $H_{XC}'$ commute with one another. 

To see why $H_{XC}'$ is equivalent to $H_{XC}$, we work in the ground state subspace of the $h$-term, where we have the additional local constraint $X^a_l X^b_l = 1$ on each link. (Again, $a \neq b$ and $\uva, \uvb \perp l$.) We thus have a single effective qubit on each link, with Pauli operators
\begin{equation}
X^{{\rm eff}}_l = X^c_l X^a_l = X^c_l X^b_l
\end{equation}
and
\begin{equation}
Z^{{\rm eff}}_l = Z^c_l = Z^a_l Z^b_l \text{,}
\end{equation}
where $\uvc \prl l$. The remaining $g$- and $K$-terms of $H_{XC}'$ are easily expressed in terms of $X^{{\rm eff}}_l$ and $Z^{{\rm eff}}_l$, and reduce to $H_{XC}$.

As in the case of fracton dipole condensation, in order to condense lineon dipoles while keeping single lineons gapped, it is convenient first to map $H'_{XC}$ to a generalized gauge theory. The gauge theory is constructed by placing three qubits on each vertex and on each link of the lattice. We first discuss the site variables, whose Pauli operators are denoted $\sfx^a_i,\sfz^a_i$ on each vertex $i$, and $a = x,y,z$. We will see that $\sfz^a_i$ creates an $\mfe^a$ lineon at the vertex $i$. In accordance with the lineon fusion rules, we impose the constraint 
\be \label{threez_matter} \sfz^x_i \sfz^y_i\sfz^z_i=1\ee 
at each site. 

On each link, the Hilbert space is the same as in the construction of the ungauged model $H'_{XC}$, except that we denote the Pauli operators for the three qubits by $\tX_l^a, \tZ^a_l$. We impose the same constraint on each link, namely $\tZ^x_l \tZ^y_l \tZ^z_l = 1$. We define operators $\tA^{ab}_i$ in terms of the $\tX_l$ by the same formula as \eqref{eqn:Aab}. 

To reduce down to the original number of degrees of freedom, we then impose, for each $a\neq b$, the Gauss law constraints 
\bea \label{initial_gauss} 
\sfx_i^a \sfx_i^b = \tA_i^{ab}.\eea 
Note that this is compatible with the constraint \eqref{threez_matter}. Moreover, only two of the three equations in \eqref{initial_gauss} are independent, because taking the product of all three equations gives the triviality $1 = 1$.

The mapping is established by the following dictionary between the operators in $H'_{XC}$ and those in the generalized gauge theory:
\begin{eqnarray}
X^a_l X^b_l &\mapsto& \tX^a_l \tX^b_l \qquad (\text{any pair } a \neq b) \\
Z^a_{\ij} &\mapsto& \sfz^a_i \tZ^a_{\ij} \sfz^a_j \qquad (\uva \prl \ij) \\
Z^b_{\ij} &\mapsto& \sfz^c_i \tZ^b_{\ij} \sfz^c_j \qquad (\uvb, \uvc \perp \ij, b \neq c) \text{.}
\end{eqnarray}
Under this mapping, the modified X-cube Hamiltonian $H'_{XC}$ becomes
\begin{equation}
\widetilde{H}'_{XC} = - g\sum_i \sum_{a\neq b} \sfx^a_i \sfx^b_i - K \sum_c \prod_{l\in c} \tZ^l_l
- h \sum_l \tX^a_l \tX^b_l \text{.}
\end{equation}

The operator dictionary tells us that, on a link $\ij \prl \uva$, the non-gauge-invariant operator $\tZ^b_\ij$ creates an excitation in the same superselection sector as the $\mfd^z_\mfe$ lineon dipole $\sfz^c_i \sfz^c_j$, where $\uvb, \uvc \perp \ij$ and $b\neq c$. This is so because the gauge-invariant local operator $\sfz^c_i \tZ^b_{\ij} \sfz^c_j$ necessarily creates a trivial, locally createable excitation. We also observe that acting with $\tZ^b_\ij$ violates the $h$-term in $\widetilde{H}'_{XC}$, and no other terms, so that $h$ can be understood as setting the energy gap for lineon dipoles. Since $g$ controls the gap for single lineons, we can control the single-lineon and lineon-dipole gaps independently in this formulation, as desired. We note that, by the same reasoning as above, acting with $\tZ^a_\ij$ ($\uva \prl \ij$) creates an excitation in the same sector as a pair of $\mfe^a$ lineons separated in the $\uva$-direction; however, as discussed in Sec.~\ref{sec:preliminaries}, such a pair is a trivial excitation. 

\begin{figure}
    \centering
    \includegraphics{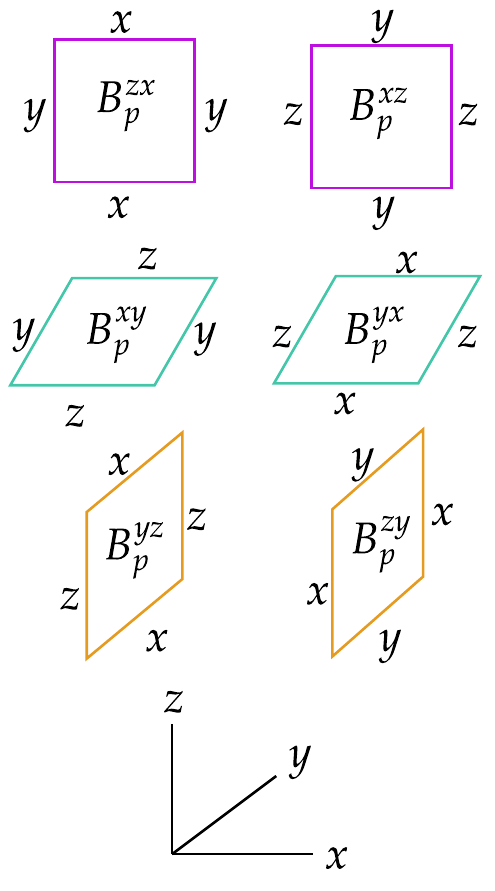}
    \caption{An illustration of the operators $B^{ab}_p$. The indices on each link $l$ denote the corresponding $\wt Z^a_l$ operators.}
    \label{fig:bab_operators}
\end{figure}

Now we introduce operators that, when added to the Hamiltonian, will play the role of kinetic energy for the lineon dipoles. Recall that a lineon dipole is a planon, moving in the plane perpendicular to its dipole moment. On each plaquette $p$ lying in the $ab$-plane, we define two operators, $B^{ab}_p$ and $B^{ba}_p$. $B^{ab}_p$ hops a $\mfd^a_\mfe$ dipole along $\uvb$, where the dipole consists of a pair of $\mfe^b$ excitations. Similarly, $B^{ba}_p$ hops a $\mfd^b_\mfe$ (consisting of a pair of $\mfe^a$'s) along $\uva$. More concretely, we define
\begin{eqnarray}
B^{zx}_p &=& \tZ^y_{l_1} \tZ^x_{l_3} \tZ^x_{l_4} \tZ^y_{l_2} \\
B^{xz}_p &=& \tZ^y_{l_3} \tZ^z_{l_1} \tZ^z_{l_2} \tZ^y_{l_4} \\
B^{zy}_p &=& \tZ^x_{l_1} \tZ^y_{l_3} \tZ^y_{l_4} \tZ^x_{l_2} \\
B^{yz}_p &=& \tZ^x_{l_3} \tZ^z_{l_1} \tZ^z_{l_2} \tZ^x_{l_4} \\
B^{xy}_p &=& \tZ^z_{l_1} \tZ^y_{l_3} \tZ^y_{l_4} \tZ^z_{l_2} \\
B^{yx}_p &=& \tZ^z_{l_3} \tZ^x_{l_1} \tZ^x_{l_2} \tZ^z_{l_4} \text{.}
\end{eqnarray}
These operators are displayed graphically in Fig. \ref{fig:bab_operators}. 
The indexing of links in each plaquette is exactly as illustrated for dual lattice plaquettes in Fig.~\ref{fig:std_labeling}. These operators are easily checked to be gauge-invariant. In each term, the two factors of $\tZ^a_l$ written on the outside of the product create/destroy lineon dipoles, while the two factors of $\tZ^a_l$ written on the inside are needed to make the term gauge-invariant. We see that for $\uva \prl l$, $\tZ^a_l$ plays the role of a gauge field, while for $\uva \perp l$, $\tZ^a_l$ is a matter field creating a lineon dipole. By taking products for instance of $B^{zx}_p$ and $B^{zy}_p$ operators, we can obtain string operators for a $\mfd^z_\mfe$ dipole moving along an arbitrary path in the $xy$-plane. Whenever the dipole turns a corner, the string operator contains a factor of $\tZ^z_l$ with $l \prl \uvz$, corresponding to the two constituent particles converting from $\mfe^x$'s to $\mfe^y$'s (or vice versa) by exchanging an $\mfe^z$. 

The following solvable Hamiltonian condenses all species of lineon dipoles, while keeping single lineons gapped: %
\begin{equation}
H_{con} = - g\sum_i \sum_{a\neq b} \sfx^a_i \sfx^b_i - \lambda \sum_{a\neq b} \sum_{p\prl \uva\times\uvb} (B^{ab}_p + B^{ba}_p ) \text{.} \label{eqn:hcon-lineon-dipoles}
\end{equation}
Here, in the last term, the sumis over plaquettes $p$ lying in the $ab$-plane. In addition to adding the kinetic energy term to $\widetilde{H}'_{XC}$, we have set $h = 0$ to allow the lineon dipoles to condense and obtain a solvable Hamiltonian. Moreover, we have also set $K = 0$, as the $K$-term can be obtained as a product of the $B^{ab}_p$ kinetic energy operators. 

It turns out that $H_{con}$ is the $\zt^2$ gauge theory we expected to find. To see this, we first recall that given the constraint $\tZ^x_l \tZ^y_l \tZ^z_l = 1$, we effectively have two qubits on each link. We define $Z^{1,2}_l, X^{1,2}_l$ Pauli operators for these two qubits, where the upper index will label the two $\zt$ gauge theories. Specifically we choose
\begin{align}
&Z^1_l = \tZ^z_l ,  &Z^2_l = \tZ^y_l \qquad  &(l \prl \uvx) \\
&Z^1_l = \tZ^y_l ,  &Z^2_l = \tZ^x_l  \qquad &(l \prl \uvy)  \\
&Z^1_l = \tZ^x_l ,  &Z^2_l = \tZ^z_l  \qquad &(l \prl \uvz) \text{.}
\end{align}
From this choice it follows that $X^1_l = \tX^z_l \tX^x_l$ and $X^2_l = \tX^y_l \tX^x_l$ for $l \prl \uvx$, with similar choices for the two other orientations of $l$. For the lineon matter fields, we define
\begin{align}
\sfx^1_i = \sfx^x_i \sfx^y_i \qquad & \sfx^2_i = \sfx^x_i \sfx^z_i \\
\sfz^1_i = \sfz^y_i \qquad & \sfz^2_i = \sfz^z_i \text{.}
\end{align}

With these choices the Gauss law equations for $ab = xy$ and $ab = xz$ become
\begin{eqnarray}
\tA^{xy}_i &=& A^1_i = \sfx^1_i \\
\tA^{xz}_i &=& A^2_i = \sfx^2_i \text{,}
\end{eqnarray}
where $A^{1,2}_i = \prod_{l \ni i} X^{1,2}_l$ is a product over the six links touching $i$. Similarly defining $B^{1,2}_p = \prod_{l \in p} Z^{1,2}_l$, the Hamiltonian becomes
\begin{eqnarray}
H_{con} &=& -  g\sum_i (\sfx^1_i + \sfx^2_i )  - \lambda \sum_{p \prl \uvx} ( B^1_p + B^2_p ) \nonumber \\ 
&-& \lambda \sum_{p \prl \uvy} ( B^2_p +  B^1_p B^2_p ) - \lambda \sum_{p \prl \uvz} (B^1_p + B^1_p B^2_p ) \text{.}
\end{eqnarray}
This is not quite the standard Hamiltonian for two decoupled $\zt$ gauge theories, due to the presence of the $B^1_p B^2_p$ terms. However, acting on the ground state we clearly have $B^1_p = B^2_p = 1$, so the ground state is the same as that of the standard Hamiltonian. Moreover, it is trivial to deform this Hamiltonian to the standard one by the obvious linear interpolation, so $H_{con}$ is indeed in the same phase as the standard deconfined $\zt^2$ gauge theory.

\ss{Condensing a subset of lineon dipoles} 

We can similarly consider what happens when only a subset of the lineon dipoles condense. First consider condensing only a single orientation of dipole, say with dipole moment along $\uvz$. 
To treat this case, we modify \eqref{eqn:hcon-lineon-dipoles} as follows:
\begin{eqnarray}
H_{con;z} &=& - g\sum_i \sum_{a\neq b} \sfx^a_i \sfx^b_i - \lambda \sum_{p\prl \uvz\times\uvx} B^{zx}_p
- \lambda \sum_{p\prl \uvz\times\uvy} B^{zy}_p \nonumber \\
&-& h \sum_{l \prl \uvx} \tX^y_l \tX^z_l - h \sum_{l \prl \uvy} \tX^x_l \tX^z_l \text{.}
\end{eqnarray}
Here we include the kinetic energy terms only for $\mfd^z_\mfe$ dipoles, and retain the $h$-terms on $x$- and $y$-links, which keep $\mfd^{x,y}_\mfe$ dipoles gapped. The $K$-term is not included as it can be obtained as a product of kinetic energy terms. A similar analysis to that above shows that, in the ground state subspace of the $h$-term, $H_{con;z}$ is precisely two independent stacks of 2d $\zt$ gauge theories, stacked along the $\uvx$ and $\uvy$ directions.

Now consider condensing dipoles with moments along both $\uvz$ and $\uvx$. 
In this case we consider 
\begin{eqnarray}
H_{con;z} &=& - g\sum_i \sum_{a\neq b} \sfx^a_i \sfx^b_i - \lambda \sum_{p\prl \uvz\times\uvx} (B^{zx}_p + B^{xz}_p)  \\
&-& \lambda \sum_{p\prl \uvz\times\uvy} B^{zy}_p 
- \lambda \sum_{p \prl \uvx \times \uvy} B^{xy}_p -  h \sum_{l \prl \uvy} \tX^x_l \tX^z_l \text{.} \nonumber
\end{eqnarray}
This includes kinetic energy terms for $\mfd^{x,z}_\mfe$ dipoles, while keeping $\mfd^y_\mfe$ dipoles gapped. Again, a similar analysis shows that this model is in the same phase as a 3d $\zt$ gauge theory plus a stack along $\uvy$ of 2d $\zt$ gauge theories, as expected.

The generalization of these constructions to the $\zn$ case is straightforward, and can be done in much the same manner as described in appendix \ref{sec:frac_dip_zn} for the case of fracton dipoles. All the prescriptions for adding Hermitian conjugates to operators are the same as in that case, and as such we will omit the details.

\ss{Phase transitions} 

{Because we are again condensing excitations that can move in 2d planes, with the number of excitations in each plane separately conserved modulo $N$ (for local processes),} the critical points to analyze are the same as those that appeared in section \ref{sec:2dphase_transitions_fracton}. The stability analysis of the critical points is identical to the previous fracton dipole case, and so we will not repeat ourselves.

\section{Lineon condensation \label{sec:1dexps}}

In this section we will discuss what happens when we condense one or more species of the lineons $\mfe^{x,y,z}$. We will see that when we condense a single species of lineon the resulting phase is a stack of deconfined 2d $\zn$ gauge theories, while if more species are condensed the resulting phase is trivial. {If $N > 4$, we find that the system enters an intermediate gapless phase upon condensation of lineons, described as an array of strongly-coupled Luttinger liquids. The transition from the XC phase into the gapless phase is expected to be of Berezinskii-Kosterlitz-Thouless (BKT) type.}

\ss{Condensed phases and dual Hamiltonians \label{sec:oned_dual_hams}}

Before getting into details let us discuss our expectations for the condensed phases. 
Consider first condensing a single type of lineon, for example $\mfe^z$. 
In the electric sector, the fusion rule $\mfe^x\mfe^y\mfe^z=1$ means that the condensed phase contains a single type of electric \ex $\mfe^x\sim (\mfe^y)^\da$, which is able to move in both the $x$ and $y$ directions by fusing with the $\mfe^z$ condensate. 
{In the magnetic sector, the $\mfe^z$ condensate leads to confinement of fractons. Consider first fracton membrane operators lying in the $xy$-plane. Such operators do not commute with string operators of the condensate $\mfe^z$ lineons, and thus acquire a surface tension. Naively, $xz$- and $yz$-plane fracton membrane operators do not acquire any tension, and indeed they do not feel the $\mfe^z$ condensate. However, in the presence of the condensate, the edges of such membranes running along the $z$-direction become locally detectable, via processes that create \emph{e.g.} a pair of $\mfe^x$ lineons, and move one of them in a small loop encircling the edge before annihilating the lineons to vacuum {(the movement around the loop creates no additional excitations due to the presence of the $\mfe^z$ condensate).} Nothing prevents adding terms to the Hamiltonian that effect such processes, giving a line tension to the $z$-direction edges of the $xz$- and $yz$-membrane operators.}
These effects result in confinement of $\mfd^{x,y}_\mfm$ dipoles, while leaving $\mfd^z_\mfm$ dipoles  deconfined.
The excitation content is thus the same as a stack of 2d $\zn$ gauge theories oriented along the $\uvz$ direction. Indeed, we will see momentarily that this is the correct identification. 

When we further condense a second species of lineon, all of the lineons become condensed {due to the fusion rule $\mfe^x\mfe^y\mfe^z=1$}, and consequently the magnetic sector is entirely confined. We therefore obtain a trivial paramagnet.

\begin{figure}
	\includegraphics{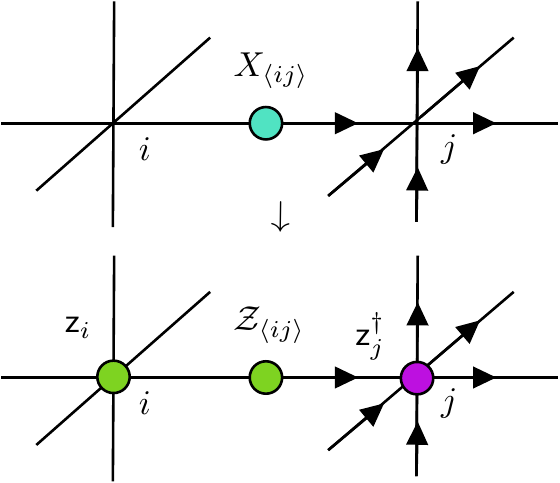}
	\caption{\label{link_gauge} The dual representation of the operator $X_\lij$ in terms of a gauge field $\mcz_\lij$ and two matter spins $\sfz_i,\sfz_j$. The arrows denote the orientations of their parent links. } 
\end{figure}

\begin{figure}
    \centering
    \includegraphics{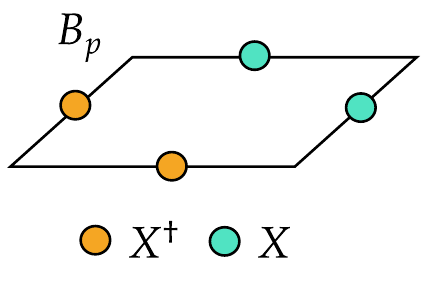}
    \caption{An illustration of the operator $B_p$ appearing in $H_{con;z}$ of \eqref{eqn:hcon-ez}.}
    \label{fig:bxy}
\end{figure}

Now we will see how these expectations are borne out at the level of explicit Hamiltonians. To condense lineons we may simply add the kinetic energy term
\be H_\lambda =- \sum_{a} \lambda^a \sum_{l\prl \uva} (X_l+X^\da_l).\ee 
To identify the phase that results upon condensing $\mfe^z$ lineons, we take only $\lambda^z \neq 0$ with $\lambda^x = \lambda^y = 0$. Moreover, we drop the $A^x_i$ and $A^y_i$ terms in $H_{XC}$, which tend to gap out $\mfe^z$ lineons and do not commute with the $\lambda^z$ kinetic energy term. We thus arrive at
\begin{eqnarray}
H_{con;z} &=& - g \sum_i A^z_i - K \sum_c B_c - \lambda^z \sum_{l \prl \uvz} X_l  \nonumber \\ &-& K' \sum_{p \prl \uvz} B_p + h.c. \label{eqn:hcon-ez}
\end{eqnarray}
Here we have also added a term with coefficient $K'$, where $B_p$ is a product of $X_l$ and $X_l^\dagger$ over the four links in the $xy$-plane plaquette $p$, as shown in Fig. \ref{fig:bxy}.
This term corresponds precisely to a process where a pair of $\mfe^x$ lineons are created, one is moved around a small $xy$-plane loop, and then the pair is annihilated. As discussed above, such a process detects the edges of fracton membrane opeators parallel to $\uvz$, and so the $K'$ term gives these edges a line tension. Moreover, if the $K'$-term is not added, then an undesirable unstable ground state degeneracy remains.

To analyze $H_{con;z}$, we first observe that all the terms included commute with one another. We can then set $X_l = 1$ for $l \prl \uvz$, upon which $B_c = B_p B_{p'}^\dagger$, where $p$ (resp. $p'$) is the bottom (resp. top) face of $c$. We see that the $K$-term is redundant and can be dropped in favor of the $K'$-term, giving a stack of $\zn$ toric codes as expected.

Now suppose we condense both $\mfe^z$ and $\mfe^x$ lineons. This is achieved by the following Hamiltonian:
\begin{equation}
H_{con;zx} = -K \sum_c B_c - \lambda^z \sum_{l \prl \uvz} X_l -  \lambda^x \sum_{l \prl \uvx} X_l \text{.}
\end{equation}
Here we have completely dropped the $g$-term to allow both species of lineons to condense. Setting $X_l = 1$ for $l \prl \uvx, \uvz$ reduces $B_c$ to a product of $X_l$ over the four links along $\uvy$. The resulting model is thus a stack of 2d Ising plaquette models (see \emph{e.g.} Ref.~\onlinecite{vijay2016fracton}) along the $\uvy$-direction. There is a ground state degeneracy associated with breaking of subsystem symmetries, but since we are not imposing these symmetries microscopically, the degeneracy is not robust.  Indeed, we can resolve the degeneracy by adding the term $-\lambda^y \sum_{l \prl \uvy} (X_l + h.c)$, which results in a trivial gapped phase with a unique ground state. Not coincidentally, this term is the kinetic energy for $\mfe^y$ lineons, which are automatically condensed, given the fusion rule $\mfe^x \mfe^y \mfe^z = 1$.

To discuss the lineon condensation phase transitions, it is convenient to work in a dual description in terms of $\zn$ spins more directly associated to the lineon excitations, following a logic similar to that used when discussing condensation of fracton and lineon dipoles. We will begin by fractionalizing $X_\lij$ in terms of dual spins that create $\mfe^a$ lineons.\footnote{See Ref. \onlinecite{pretko2020odd} for a similar type of duality.} We do this by writing, for each link $\lij \prl \uva$, 
\be \label{lineonduality} X_\lij = \sfz^a_i \mcz_{\lij}(\sfz_j^a)^\da , \qquad Z_{\lan ij\ran} = \mcx_\lij, \ee 
where $\sfz^a_i$ is to be thought of as creating an $\mfe^a$ lineon on site $i$, and where the $\mcz_\lij,\mcx_\lij$ are dual gauge field variables. 
The dual representation of $X_\lij$ is shown in figure \ref{link_gauge}. 
As in section \ref{sec:2d_lineons}, in accordance with the lineon fusion rules, at each site we impose the constraint 
\be \label{lineonthreez} \sfz_i^x \sfz_i^y\sfz_i^z = 1.\ee 

The dual representation and the constraint \eqref{lineonthreez} are invariant under the $\zn$ gauge transformations 
\bea \label{gauge_tforms} \sfz^a_i & \mt \zeta_i^b ( \zeta_i^c)^* \sfz^a_i, \\ \mcz_{\lan ij \ran \prl \uva } & \mt  (\zeta_i^b)^* \zeta_i^c \mcz_{\lan ij \ran}\zeta_j^b(\zeta_j^c)^*,\eea
where the $\zeta_i^a$ are valued in the $N$th roots of unity and $a,b,c$ are all distinct. The Gauss laws originating from this gauge redundancy are, for each $a\neq b$, 
\be \label{onedgauss} \sfx^a_i (\sfx^b_i)^\da  = \mcx_{\lan i, i+\uva\ran } \mcx^\da_{\lan i, i-\uva\ran} \mcx_{\lan i,i+\uvb\ran}^\da\mcx_{\lan i,i-\uvb\ran},\ee 
where the terms on the RHS are precisely the duals of the $A^a_i$ operators appearing in $H_{XC}$ (c.f. figure \ref{fig:xc_ham}).

There are two types of gauge-invariant operators that create lineons which will be important in what follows. The first is the dual representation of the $X_\lij$ operators, which hop $\mfe^a$ lineons along the $\uva$ direction. The next-simplest term is a ``ring-exchange'' operator
\bea \label{ring_exchange} R_i^a & = \sfz^a_i \mcz_{\lan i,i+\uvb\ran } (\sfz^a_{i+\uvb})^\da \mcz_{\lan i+\uvb,i+\uvb+\uvc\ran }^\da \sfz^a_{i+\uvb+\uvc}  \\ & \qquad \times \mcz^\da_{\lan i + \uvc,i+\uvb+\uvc\ran} (\sfz^a_{i+\uvc})^\da \mcz_{\lan i,i+\uvc\ran}, \eea
where as usual $a,b,c$ are all distinct. This operator creates four $\mfe^a$ lineons at the corners of a plaquette normal to $\uva$ and with a corner at the site $i$, and will be seen to play an important role in our analysis of the condensation transition.

The Hamiltonian can now be written as 
\bea \label{nonfixed} H_{con} &= -g\sum_{a\neq b} (\sfx_i^a)^\da \sfx_i^b - K\sum_c  \mcb_c \\ & 
\qquad - \sum_a \lambda^a \sum_{\lij \prl \uva} \sfz^a_i \mcz_\lij (\sfz^a_j)^\da + h.c, 
\eea
where $\mcb_c$ is $B_c$ but with $\mcz_\lij$ operators instead of $X_\lij$ operators. Magnetic excitations are gapped, so as in the previous examples we can ignore fluctuations of the $\mcz_\lij$ generalized gauge field, which indeed are not present in \eqref{nonfixed}.

{If we set $\mcz_\lij = 1$, we obtain an effective matter theory for the lineons, which is valid when considering correlation functions of local, gauge-invariant operators. This effective matter theory enjoys planar $\zn$ subsystem symmetries, arising from the planar $\zn$ conservation  laws for lineons in the X-cube phase discussed in Sec.~\ref{sec:preliminaries}. Concretely, the Hamiltonian and any perturbations are required to be invariant under the transformation}
\begin{equation}
\sfz^a_i \to \zeta^b_i \zeta^c_i \sfz^a_i \text{,}
\end{equation}
{where $a,b,c$ are all distinct, and where $\zeta^b_i$ and $\zeta^c_i$ are constant within planes normal to $\uvb$ and $\uvc$, respectively.}

\subsection{Phase transitions for $N\leq 4$ \label{sec:oned_critical_smalln}}

We now examine the nature of the phase transitions that occur when the lineons condense. We will first consider only the simplest case when a single species of lineon condenses, which as in the previous section we will take to be $\mfe^z$. We will therefore take $\lambda^a= (0,0,\lambda)$, and consider the condensation transition that occurs as a function of $g/\lambda$. 

In this limit, $\sfx^x,\sfx^y$ both commute with the Hamiltonian \eqref{nonfixed}, and to study the phase transition we may work in the $\sfx^x = \sfx^y = 1$ subspace. In this subspace the Hamiltonian \eqref{nonfixed} clearly reduces to an array of 1d $\zn$ clock model wires oriented along the $\uvz$ direction, all of which are coupled to the (non-fluctuating) $\zn$ gauge field $\mcz_\lij$. In this limit, the critical point is simply described by a decoupled array of 1d critical clock models, with the critical fluctuations occurring on each wire independently. 

Of course the limit where the $\zn$ chains are completely decoupled is very fine-tuned, and we are interested in the fate of the decoupled fixed point under generic perturbations respecting the effective subsystem symmetry. This can be determined by performing an RG analysis -- if all of the perturbations which couple different chains are irrelevant, the decoupled fixed point will correctly describe the phase transition in a finite region of parameter space. If some of the perturbations are relevant, they may drive the system into a nontrivial coupled fixed point, or may lead to the transition being driven first order. Using our knowledge of the critical properties of the 1d $\zn$ clock model, the RG analysis can be performed straightforwardly. Two classes of terms need to be considered, those which are a) invariant under separate $\zn$ transformations on each wire or are b) built from the ring exchange terms of \eqref{ring_exchange}.

\sss{$N=2$}

First, consider the case where $N=2$. The decoupled critical point is then described by an array of critical Ising CFTs. Energy-energy interactions are an important class of couplings between pairs of chains. Labeling the chains by $w$ and assuming translation symmetry, these couplings give us a perturbation to the decoupled fixed point which can be written as 
\be \label{ee_coupling} \d S_\ep = \sum_{w\neq w'} C_{w-w'} \int dz\, d\tau\, \ep_w\ep_{w'}, \ee 
where $\ep_w$ is the energy operator on the chain $w$ (note that since there is no $\ep^2$ operator in the 1d Ising CFT, we only sum over $w'\neq w$).
Since the critical fluctuations are only in the $z$-$\tau$ spacetime plane, the relevance of $\d S$ is obtained by comparing the scaling dimension of $\ep_{w} \ep_{w'}$ to 2. 
Since the dimension of the energy operator in the 1d Ising model is 1, the couplings $C_{w,w'}$ are all marginal at tree level. %

However, it is unnecessary to actually compute the beta functions for the $C_{w-w'}$ couplings. This is because there is another gauge-invariant coupling which is relevant, which as we will see  destabilizes all of the decoupled critical points when $N\leq 4$. This is the ring-exchange term \eqref{ring_exchange}, which at the critical point is written as (dropping the $\mcz$ gauge fields as they do not affect calculations of scaling dimensions at the critical point)
\bea \d S_r & = r\sum_{i,a} \int \int dz\, d\tau\, R_i^a  \\ 
 & \ra  r\sum_{w_{1,\dots,4} \in \sq}  \int dz\, d\tau\, \s_{w_1} \s_{w_2} \s_{w_3} \s_{w_4},\eea 
where $\s_w$ is the spin operator on wire $w$ and the notation $w_{1,\dots,4}\in \sq$ means a sum over configurations of four wires $w_{1,\dots,4}$ whose $x,y$ coordinates form the corners of a square with unit normal along $\uvz$. Since the scaling dimension of $\s$ is $\De_\s = 1/8$, the scaling dimension of the above term is 
\be \De_r = 4 \De_\s = \frac12 < 2,\ee 
which is very relevant. Higher-order contributions to the beta function for $r$ do not lead to fixed points accessible at small couplings. We therefore conclude that the transition is likely generically first-order, in agreement with suggestions from numerics.\cite{slagle2017fracton}
The same conclusion holds when mutiple species of lineons condense simultaneously.

\sss{$N=3$}

Now consider $N=3$. In this case the putative decoupled critical point corresponds to an array of $\zz_3$ clock models.
At the critical point of the $\zz_3$ chain, the scaling dimension of the spin operator $\s$ is $\De_\s = 2/15$, while that of the energy operator is $\De_\ep = 4/5$.\cite{dotsenko1984critical} Therefore both the ring-exchange term and the energy-energy couplings are relevant, and the transitions are again expected to be first order. 

\sss{$N=4$}

Next up is $N=4$. 
While in this case there is a one-parameter critical line of fixed points (the Ashkin-Teller line), the scaling dimension of the spin operator on each chain is always $\De_\s = 1/8$.\cite{delfino2004universal} As such the ring-exchange term is again relevant, and destabilizes the decoupled critical point.

\ss{Phase transitions for $N>4$ \label{sec:oned_critical_largen}}

For $N > 4$, the 1d chains are more conveniently dealt with using a continuum XY description.\cite{jose1977renormalization} We do this by writing the $\sfz^z$ and $\sfx^z$ variables as exponentials of slowly-varying fields $\cp_w(z)$ and $\ct_w(z)$, so that in the IR near the phase transition we have the approximate identifications 
\bea  \label{irspins}  \sfz^z_{(w,z)}  & \sim \exp\(i\cp_w(z)\), \\  \sfx^z_{(w,z)}  &\sim   \exp\( i \frac1N \p_z\ct_w(z)\) = \exp\(i \frac\twp N \pi_{\Phi_w}(z)\),\eea
where $\pi_{\cp_w} = \frac1\twp \p_z \ct_w$ is the momentum conjugate to $\cp_w$, and where a suitable continuum limit is taken in the $\uvz$ direction, assuming a continuous transition. 
The $\zn$ nature of the problem can be accounted for by including a $\zn$ anisotropy term $\cos(N\cp_w)$ in the action.

The most general action for the putative fixed point can then be written as $S = S_0 + S_{I}$, where $S_I$ contains interactions and where the first term represents a non-interacting quadratic fixed point:
\be \ba \label{free_chains} 
S_0 & = \frac1{\fpi} \sum_{w,w'} \int  dz\, d\tau  \, \big( 2 \d_{w,w'} \p_\tau\cp_w\p_z\ct_{w'}  \\ &\qquad  + g_{w-w'}\p_z\ct_w\p_z \ct_{w'} + \eta_{w-w'}\p_z\cp_w \p_z \cp_{w'} \big),  \ea \ee 
where we have assumed translation invariance for the interchain derivative couplings $g_{w-w'}, \eta_{w-w'}$. 

$S_I$ contains the $\zn$ anisotropy and further cosines in $\cp_w,\ct_w$ allowed by gauge invariance, with the most important ones for the present problem being  
\bea S_I & = a^{-2}\int dz\, d\tau\,\Big[u \sum_w \cos(N\cp_w)  \\ & \qquad +   r\sum_{w_1, \dots, w_4 \in \sq }    \cos\Big( \cp_{w_1} - \cp_{w_2} + \cp_{w_3} - \cp_{w_4}\Big)  \\ & \qquad + s \sum_w \cos(\ct_w) %
+ \cdots\Big],\eea 
where $a$ is the lattice spacing.
Despite the fact that we naively need $u\ra\infty$ to enforce the $\zn$ nature of the problem, $u$ can in fact be treated perturbatively.\footnote{This can be shown by requiring that the known physics of the XY model be recovered in the $N\ra\infty$ limit; see Refs. \onlinecite{elitzur1979phase,frohlich1981kosterlitz}.}

\sss{Decoupled limit} 

 Let us first compute the scaling dimensions of the terms in $S_I$ in the case where there are no interchain derivative couplings, i.e. let us first consider setting 
 \be g_{w-w'} = g\d_{w,w'} ,\qquad \eta_{w-w'} = \eta\d_{w,w'}.\ee 
  The scaling dimensions of the most relevant gauge-invariant cosines are then found to be  
\bea \label{deltas} \De_u & = \frac{N^2}{2R^2}\\ \De_{r} & = \frac{2 }{R^2} \\ 
 \De_s & = \frac{R^2}{2}
  ,\eea
where we have defined 
\be R^2 \equiv \sqrt{\eta/g}.\ee   

The relevance of a given term in $S_I$ is determined by comparing the appropriate scaling dimension to two. 
When the field term is small, so that the lineons are not condensed, $R^2$ is also small. Here the $\cos(\ct_w)$ operators are relevant, pinning the values of the $\ct_w$ fields. This regime occurs for $R^2 < 4$. In the limit of large field where the lineons are condensed, $R^2$ is large, and here the $\cos(N\cp_w)$ term is relevant. This happens for $R^2> N^2/4$. In the absence of the ring exchange term this would give an intermediate massless regime for $4 < R^2 <N^2/4$ with no relevant perturbations. However, we see that the ring exchange term is relevant for all $R^2 > 1$, meaning that there is in fact no choice of $R^2$ for which all three terms in \eqref{deltas} are irrelevant. 

It remains to understand the nature of the intermediate $R^2$ regime when the ring exchange term is the dominant relevant perturbation. What happens in this case is explained in appendix \ref{app:ring_exchange}. In brief, we find that the ring-exchange term leads to a putative fixed point corresponding to an anisotropic version of the exciton Bose liquid,\cite{paramekanti2002ring,you2020fracton,seiberg2020exoticI,seiberg2020exoticII} but that the three-dimensional nature of the problem produces an instability with respect to lineon dipole condensation. We then conclude that when the ring exchange term dominates, the transition is expected be first order.

\sss{General derivative couplings} 

That the decoupled fixed point is destabilized by the ring-exchange terms does not necessarily mean that there do not exist stable fixed points described by the quadratic action \eqref{free_chains} --- it only means that any such fixed point must owe its existence to interchain derivative couplings which frustrate ring-exchange processes. {In this section we show numerically that a relatively simple choice for the derivative couplings can be made which renders the Gaussian fixed point stable with respect to low-body finite-range cosines. Whether or not the chosen couplings ensure stability with respect to {\it all} possible deformations is a more complicated question, which we leave to future work.} {However, we also argue that there exist other, more complicated, choices of interchain couplings that result in a truly stable fixed point.}

We begin by Fourier transforming in the $x$,$y$ directions normal to the chains, writing $S_0$ as 
\bea \label{free_chains_qspace} 
S_0 & = \frac1{\fpi} \int \frac{d^2\bfq}{(\twp)^2} \,\int  dz\, d\tau  \, \Bigg[ 2 \p_\tau\cp_\bfq^*\p_z\ct_\bfq  \\ &\qquad  v_\bfq \( \frac1{R^2_\bfq} \p_z\ct_\bfq^*\p_z \ct_\bfq + R^2_\bfq \p_z\cp_\bfq^* \p_z \cp_\bfq \) \Bigg],  \eea 
where {the $q_x$ and $q_y$ momentum integrations run over $[0,2\pi)$} (with the interchain spacing set to unity), and where we have defined
\be v_\bfq \equiv \sqrt{\eta_\bfq g_\bfq},\qquad R^2_\bfq \equiv \sqrt{\eta_\bfq/g_\bfq}.\ee 
For stability we require that the couplings be chosen so that $0 < R_\bfq^2 < \infty$ for all $\bfq$; in what follows we will assume that this is the case. 

The scaling dimensions of perturbations to this fixed point can be computed as various integrals of $R^2_\bfq$-dependent functions. 
Therefore to demonstrate the existence of a stable fixed point, all that remains is to find an appropriate choice of $R^2_\bfq$ which renders all the allowed perturbations irrelevant. The problem of finding such a function $R^2_\bfq$ is very closely related to the problem of constructing examples of ``sliding Luttinger liquids'',\cite{mukhopadhyay2001sliding,vishwanath2001two} which are anisotropic metallic phases built from strongly coupled arrays of Luttinger liquids.

At the fixed point \eqref{free_chains_qspace}, general cosines of the $\cp$ and $\ct$ variables have scaling dimensions 
\bea \De_{\cos(\sum_w \a_w \cp_w)} & = \frac12 \int \frac{d^2\bfq}{(\twp)^2} \frac{|\a_\bfq|^2}{R_\bfq^2} \\ 
\De_{\cos(\sum_w \b_w \ct_w)} & = \frac12 \int \frac{d^2\bfq}{(\twp)^2} R_\bfq^2|\b_\bfq|^2, \label{eqn:general-dims}
 \eea
 where $\a_w, \b_w$ are {integer-valued functions of the wire index $w$, which are non-zero only for finitely many $w$.}  %
 
 \begin{figure}
 	\includegraphics{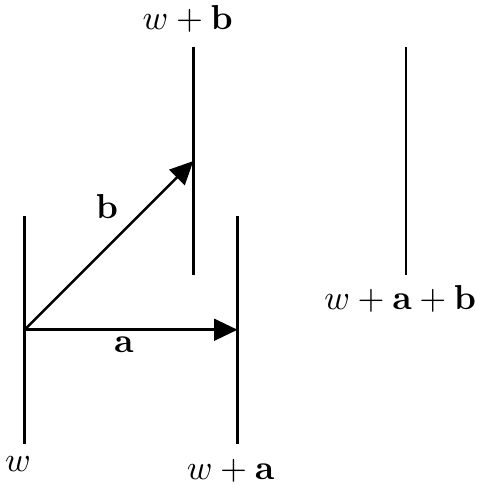}
 	\caption{\label{fig:ring_exchange} An illustration of the geometry of a ring exchange term $R^{\bfa\bfb}$. Here the vectors $\bfa,\bfb$ are in the $x$-$y$ plane and connect four wires to form a rectangle as shown.  }
 \end{figure}
 
 The simplest gauge-invariant cosines are $\cos(N\cp_w)$, $\cos(\ct_w)$, $\cos(\ct_w - \ct_{w+\bfa})$, and $\cos(R^{ab})$, where now the more general ring-exchange operator $R^{\bfa\bfb}$ is defined as (again omitting the unimportant $\mcz$ gauge field)
\be R^{\bfa\bfb} = \cp_w - \cp_{w+\bfa} + \cp_{w+\bfa+\bfb} - \cp_{w + \bfb},\ee 
where $\bfa \perp \bfb$ form the edges of a rectangle (see figure \ref{fig:ring_exchange}). Using the above formula, we see that these have the scaling dimensions 
\bea \label{qspace_sdims} \De_{\cos(N\cp_w)} & = \frac{N^2}{2R^2_\bfzero} \\ 
\De_{\cos(\ct_w)} & = \frac{R^2_\bfzero}{2} \\ 
\De_{\cos(\ct_w - \ct_{w+\bfa})} & = \frac12  \int \frac{d^2\bfq}{(\twp)^2} R^2_\bfq (1 - \cos(\bfa \cdot \bfq)) \\ 
\De_{\cos(R^{\bfa\bfb})} & = \int \frac{d^2\bfq}{(\twp)^2}\frac{ |1 - e^{-i\bfa\cdot\bfq} - e^{-i\bfb\cdot\bfq} + e^{i(\bfa+\bfb)\cdot\bfq} |^2}{2R^2_\bfq}. 
 \eea 

By taking $N$ large enough we see that we can always make the first term (the $\zn$ anisotropy) irrelevant; hence we can ignore it for the time being. For the remaining terms, we see that in order to make the above scaling dimensions large we should take $R_\bfq^2$ to be a) everywhere nonzero, b) small at the wavevectors where the numerator in the integrand of the expression for $\De_{\cos(R^{\bfa\bfb}_\cp)}$ is small, and c) large enough at other wavevectors so that the cosines of $\ct$ are kept irrelevant.  

We will not concern ourselves with a completely general analysis of which functional forms of $R^2_\bfq$ do the job, and will be content with simply demonstrating that one particular choice of $R^2_\bfq$ works. 

One such choice, which was employed in Ref. \onlinecite{mukhopadhyay2001sliding} in the context of a slightly different problem, is 
\be R^2_\bfq = \eta_0 \( 1 + \eta_1 [ \cos(q_x) + \cos(q_y)] + \eta_2 \cos(q_x)\cos(q_y) \)^2,\ee 
with $\eta_0>0$. For stability $R_\bfq^2$ must be nonzero for all $\bfq$, which ends up restricting $\eta_1,\eta_2$ to be such that 
\be -\frac12 < \eta_1 < \frac12,\qquad   2|\eta_1|-1 < \eta_2 < 1.\ee 
The dimensions of the first two operators in \eqref{qspace_sdims} are easy to compute, and are  
\bea \De_{\cos(\ct)} & = \frac{\eta_0}2 \( 1 + \eta_1^2 + \frac{\eta_2^2}4\)  ,\\  \De_{\cos(N\cp)} & = \frac{N^2}{2\eta_0(1+\eta_1^2 + \eta_2^2/4)}.\eea

\begin{figure}
	\includegraphics[width=.45\textwidth]{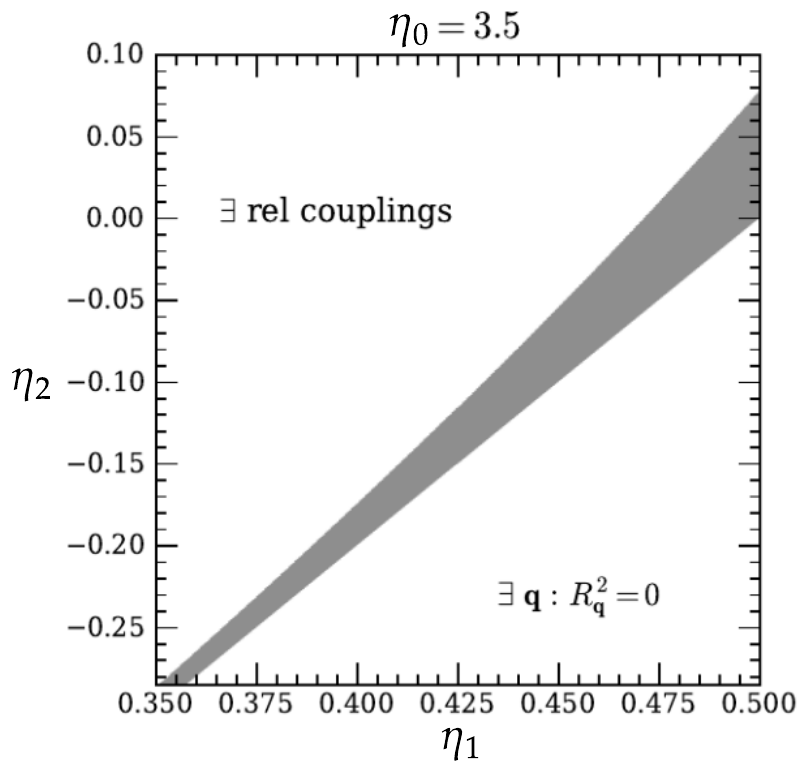}\caption{\label{fig:stability_fig} A region of stability (shaded gray area) in the $\eta_1$-$\eta_2$ plane, where the fixed-point model has no relevant perturbations. In the upper region marked ``$\exists$ rel couplings'' there exist relevant deformations to the fixed point, while in the lower region the model is unstable, as there is always some $\bfq$ for which $R^2_\bfq=0$. }
\end{figure}

To figure out if there are any regions in the $\eta_1$-$\eta_2$ plane where all of the scaling dimensions in \eqref{qspace_sdims} are greater than 2, we resort to a numerical search. {For these purposes we restrict the search to couplings involving fields up to 10th nearest neighbors.} We find small regions of stability near the boundaries of the region in which $R^2_\bfq$ is strictly positive; one such region is shown as the gray shaded area in figure \ref{fig:stability_fig}. In this region $\cos(N\cp)$ is irrelevant for all $N>4$. 
Note that although the fixed points described here are very anisotropic, they are nevertheless not completely decoupled (indeed, the coupling is crucial for their stability), unlike the fixed points considered in section \ref{sec:2dphase_transitions_fracton}.

Of course, demonstrating that the scaling dimensions in \eqref{qspace_sdims} are all greater than 2 does not actually guarantee stability, as there may be more complicated higher-body / longer-range operators which are relevant. Note that as a matter of principle it is always possible to construct a stable fixed point with sufficiently complicated interchain interactions, as was shown in Ref.~\onlinecite{plamadeala2014perfect}. {In that work, authors showed that a single quantum wire of interacting bosons has a stable gapless ``perfect metal'' phase, where all cosine operators are irrelevant, when the number of channels is $8 k$, with $k \geq 3$. The action is quadratic and crucially involves interchannel derivative couplings, as in $S_0$ of \eqref{free_chains}. It was also pointed out that arrays of such perfect metals give rise to stable gapless phases in 2d and 3d. In the present case, we can employ these results by bundling the wires of our model into groups of 24 (which breaks full translation invariance), and choosing interchain couplings within each bundle to put it in a perfect metal phase. The resulting system of decoupled bundles is then a stable gapless phase.}

Even with the present choice of relatively simple interchain couplings, any relevant cosines must necessarily involve rather high-body interactions\footnote{The scaling dimensions of higher-body interactions of the $\Theta$ fields are easy to check, since $R_\bfq^2$ only contains a handful of Fourier modes. $1/R_\bfq^2$ on the other hand contains an infinite number of Fourier modes, which makes a systematic search of the scaling dimensions of the $\Phi$ cosines more difficult.} and / or involve very long-range interchain couplings. Therefore as in the case of the ``almost perfect metals'' of Ref.~\onlinecite{murthy2020almost}, any relevant deformations are likely to have very small bare values, meaning that the examples constructed above will only show instabilities at very low temperatures.

With the aforementioned caveats about stability in mind, we thus arrive at a scenario where the XC model is separated from the condensed phase by a massless critical phase extending over a finite region of parameter space. The transition from the XC phase to the intermediate massless region is likely to be of BKT type, since the operator which becomes relevant in the XC phase is the cosine $\cos(\ct)$. The nature of the transition from the massless region to the condensed phase is not completely clear, and may depend on the type of operator which becomes relevant at the transition. 

Finally, we note that although in this section we have only considered the transitions that occur as a single species of lineon is condensed, the generalization to the case where multiple species condense simultaneously is straightforward, due to the fact that couplings between perpendicular wires are always less relevant than those between parallel wires. {In particular, when multiple lineons condense we continue to find no stable fixed points for $N\leq 4$.}

\section{Discussion \label{sec:disc}} 

In this paper we have discussed various types of condensation transitions in the $\zn$ XC model, {with the phases on both sides of the transitions described in terms of various types of gauge theories.}
We have identified continuous condensation phase transitions, {and intermediate gapless phases,} provided that $N>4$, although our analysis does not rule out the existence of continuous transitions for smaller $N$. 
Clearly there is a lot more work along these lines that may be done.

One obvious extension of the present work is to consider similar decoupled critical points in other fracton models, and to understand them using our existing knowledge of 1d and 2d critical phenomena. This strategy is likely to yield many other interesting examples.

One interesting question to ask is then whether or not there exists a continuous phase transition {between a fracton phase and a phase without fracton order}, where the critical modes at the transition fluctuate in the full four-dimensional spacetime. In fact, one such critical point was proposed in Ref. \onlinecite{vijay2017isotropic}. This work discussed the transition between the XC model and decoupled stacks of two-dimensional $\zn$ gauge theories. {Based on a duality between a model of coupled $\zn$ gauge theory layers and a 3d $\zn$ gauge theory, it was proposed that this transition is first order for $N \leq 4$, but that there may be a continuous transition out of the XC phase for $N > 4$. This transition is dual to that between the deconfined phase of 3d $\zn$ gauge theory and a massless Coulomb phase, which was claimed to be continuous in early Monte Carlo studies.}\cite{creutz1979monte} However, we believe that this transition is in fact likely weakly first order. Indeed, starting from the Coulomb phase, one may obtain the $\zn$ gauge theory by condensing electric particles with charge $N$. If we assume a second order transition, the critical point can presumably be described by a charge-$N$ Higgs field coupled to a $U(1)$ gauge field. The Higgs transition with $n\lesssim365$ flavors in 3d is however generically made first order by fluctuations,\cite{halperin1974first} and hence the transition from the XC model to the intermediate massless phase is also likely to be weakly first order. 

One possible route to a continuous transition with 3d {character} could potentially lie in Higgs transitions into the XC phase, of the type studied in Refs.\cite{ma2018fracton,bulmash2018higgs} The nonstandard dispersion of the matter fields in these examples may help to stabilize against a fluctuation-induced first-order transition, although the ultimate character of the phase transition may also end up being quasi 2d. Another possible route lies in finding a theory where the decoupled fixed point is unstable, but can be shown to flow to a stable fixed point with nontrivial couplings between different planes / chains. We leave a more detailed treatment of these possibilities to future work. 

 \section*{Acknowledgements} EL thanks Dave Aasen, Han Ma, T. Senthil, and Yizhi You for discussions. EL is supported by the Fannie and John Hertz Foundation and the NDSEG fellowship. The research of MH is supported by the U.S. Department of Energy, Office of Science, Basic Energy Sciences (BES) under Award number DE-SC0014415. This work was also partly supported by the Simons Collaboration on Ultra-Quantum Matter, which is a grant from the Simons Foundation (651440, MH).
 
 \appendix

 \section{$\zn$ fracton dipole condensation \label{sec:frac_dip_zn}}
 
 \begin{figure} \centering
 	\includegraphics{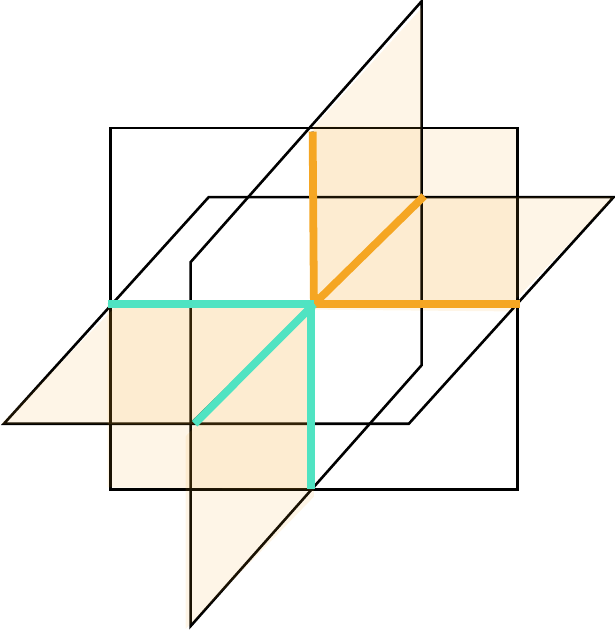}
 	\caption{\label{fig:big_gauss_law} A reference for the pattern of Hermitian conjugations appearing in the Gauss law \eqref{zn_full_gauss_law}. Links shaded in orange denote $X_\sfl^\da$ operators, and those shaded in teal denote $X_\sfl$ operators. Each orange plaquette denotes a $\mcx^\da_\sfp$ operator, and the unshaded plaquettes denote $\mcx_\sfp$ operators. } 
 \end{figure}
 
 \begin{figure} \centering
 	\includegraphics{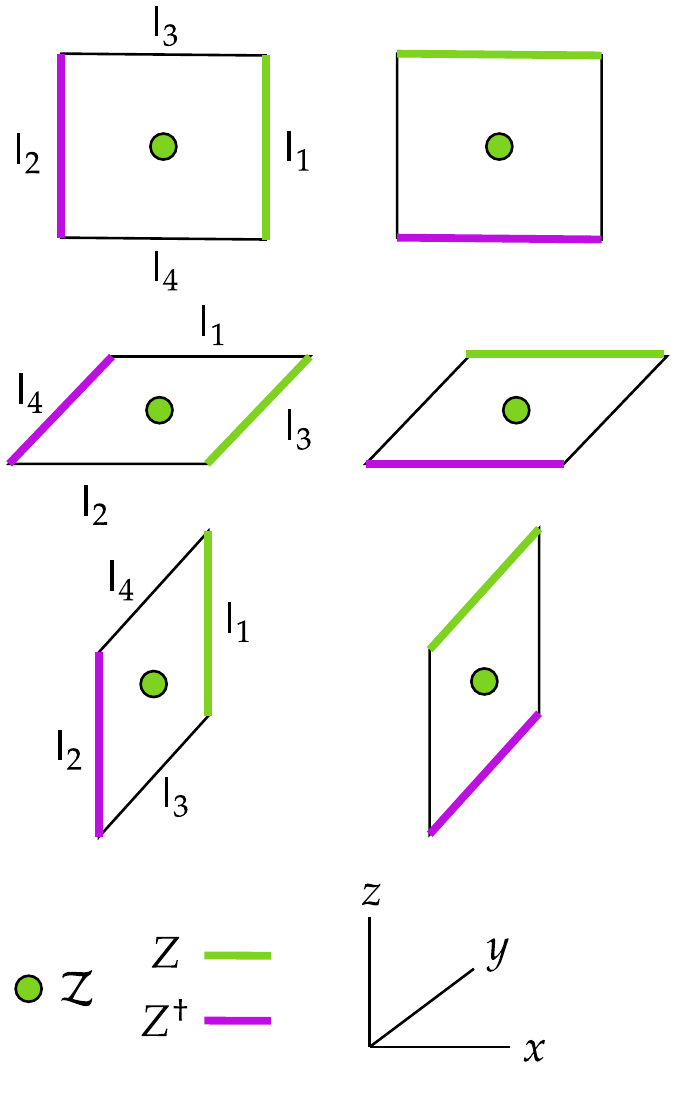}
 	\caption{\label{fig:standard_plaquettes} The kinetic terms employed when condensing dipoles (plus their Hermitian conjugates, not shown). The standard labeling of $l_{1,\dots ,4}$ for each orientation of plaquette is indicated in the left column. } 
 \end{figure}
 
 In this appendix we discuss how the generalized gauge theory employed in the discussion of fracton dipole condensation in section \ref{sec:frac_dip_phases} is generalized to the $\zn$ case. The generalization is rather straightforward, with the difficulties lying only in keeping track of the correct ways to take Hermitian conjugates in various expressions. 
 
 As in section \ref{sec:frac_dip_phases} we dualize the XC model by writing it in terms of $\zn$ matter qubits $\sfx,\sfz$ on the vertices $\sfi$ of the dual lattice, and $\zn$ gauge qubits $\mcx,\mcz$ on the plaquettes $\sfp$ of the dual lattice. 
 
 We then add additional $\zn$ qubits $X,Z$ on the links of dual lattice which represent the dipole degrees of freedom. The gauss law reads, in analogy to \eqref{z2_full_gauss}, 
 \be \label{zn_full_gauss_law} \sfx_\sfi = \prod_{\sfl,\sfp \in \p\sfi_{out}}\mcx_\sfp^\da X_\sfl^\da \prod_{\sfl',\sfp'\in \p\sfi_{in}}   \mcx_{\sfp'}  X_{\sfl'}.\ee 
 Here $\p\sfi_{out}$ consists of the collection of outward-oriented plaquettes and links neihboring $\sfi$, while $\p\sfi_{in}$ consists of the inward-oriented plaquettes and links. Here outward-oriented (inward-oriented) plaquettes are those whose centers have coordinates with respect to $\sfi$ which are all of the same sign (of different signs), and are indicated as shaded (not shaded) in figure \ref{fig:big_gauss_law}. Outward-oriented links are those oriented parallel to the coordinate axes (orange in figure \ref{fig:big_gauss_law}), and inward-oriented links are those oriented anti-parallel (cyan in figure \ref{fig:big_gauss_law}). 
 
 With this notation and the above Gauss law, the gauge-invariant kinetic terms which hop dipoles across a given plaquette are then $\mcz_\sfp Z_{\sfl_1}Z_{\sfl_2}^\da$ and $\mcz_\sfp  Z_{\sfl_3} Z_{\sfl_4}^\da$, together with their Hermitian conjugates. The correct $\zn$ generalization of the Hamiltonian \eqref{fracdip_ham} is then 
 \bea  H_{con} & = -g \sum_\sfi \sfx_\sfi - h\sum_\sfl X_\sfl - \l \sum_\sfp \mcz_\sfp (Z_{\sfl_1}Z_{\sfl_2}^\da + Z_{\sfl_3} Z_{\sfl_4}^\da ) + h.c, \eea 
 where as before we have ignored the dual representation of the $A^a_\sfi$ terms, which will not be important for discussing the condensed phase. 
 
 To demonstrate that $H_{con}$ above is equivalent to the Hamiltonian for deconfined $\zn$ gauge theory in the $h/\l \ra 0$ limit, we again need to perform a unitary transformation which decouples the plaquette degrees of freedom. This is done by directly generalizing the analysis of the $\zt$ case. 
 
 To begin, define the operators 
 \be \Pi_{\sfp k} = \frac1N \sum_{m\in \zn} (Z_{\sfl_1} Z_{\sfl_2}^\da)^m e^{ikm},\ee 
 which form a complete set of Hermitian projectors. Now define the unitaries 
 \be U_\sfp = \sum_k \Pi_{\sfp k} \mcx^k_\sfp,\ee 
 and as before let $\mcu \equiv \prod_\sfp U_\sfp$. The unitary $\mcu$ conjugates operators as 
 \bea \mcu^\da \mcx_\sfp \mcu & = \mcx_\sfp \\ \mcu^\da  \mcz_\sfp \mcu & = Z_{\sfl_1}^\da Z_{\sfl_2} \mcz_\sfp \\ 
 \mcu^\da  X_{\sfl_1} \mcu  & = \mcx_\sfp X_{\sfl_1} \\
 \mcu^\da X_{\sfl_2} \mcu & = \mcx_\sfp^\da X_{\sfl_2} \\ 
 \mcu^\da X_{\sfl_{3,4}} \mcu & = X_{\sfl_{3,4}}\\ 
 \mcu^\da Z_{\sfl_\a} \mcu & = Z_{\sfl_\a},\eea 
 which can be verified using the orthogonality and completeness of the $\Pi_{\sfp k}$ as well as the relations 
 \bea X_{\sfl_1} \Pi_{\sfp k} & = \Pi_{\sfp k-1} X_{\sfl_1} \\ 
 X_{\sfl_2}  \Pi_{\sfp k} & = \Pi_{\sfp k+1} X_{\sfl_2} \\ 
 X_{\sfl_{3,4}}\Pi_{\sfp k} & = \Pi_{\sfp k} X_{\sfl_{3,4}}.\eea 
 
 \begin{figure} \centering
 	\includegraphics{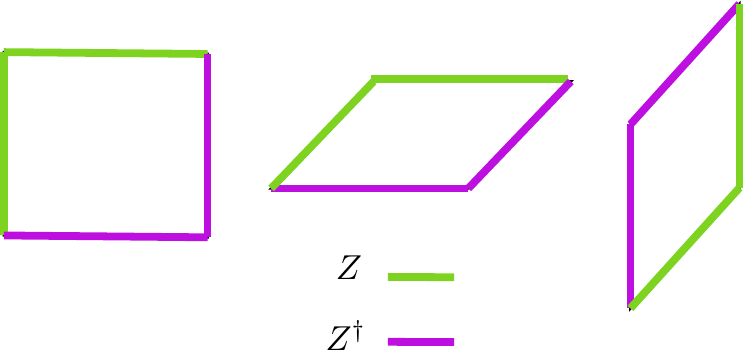}
 	\caption{\label{fig:gauge_theory_plaquettes} The plaquette terms appearing in the $\zn$ gauge theory (Hermitian conjugates not shown). } 
 \end{figure}
 
 Using these relations and referring to the pattern of Hermitian conjugation in figure \ref{fig:standard_plaquettes}, one sees that under conjugation by $\mcu$, the Gauss law constraint maps to 
 \be \sfx_\sfi = \prod_{\sfl\in \p\sfi_{out}, \sfl' \in \p\sfi_{in}} X_\sfl X_{\sfl'}^\da,\ee 
 which is the desired Gauss law of $\zn$ gauge theory. 
 
 By conjugating the kinetic term in $H$ with $\mcu$, one sees that as in the $\zt$ case, the conjugation generates a set of plaquette terms identical to that of the standard presentation of $\zn$ gauge theory, together with a trivial paramagnet coming from the plaquette variables. After getting rid of the plaquette degrees of freedom, the Hamiltonian is 
 \bea H_{con}' & = \mcu^\da H_{con} \mcu \\ & = -g \sum_\sfi \sfx_\sfi - \l \sum_\sfp B_\sfp + h.c,\eea 
 where the $B_\sfp$ terms are given by the operators appearing in figure \ref{fig:gauge_theory_plaquettes}. This completes our identification of the condensed phase with deconfined $\zn$ gauge theory.

 \section{RG flow for stacked Ising models \label{layeredIsing}}
 
 In this appendix we argue that $L$ stacked layers
 of critical 2d Ising models are always unstable 
 with respect to interlayer interactions that preserve the $\zt^L$ subsystem symmetry
 \be \zt^L : \psi_i \mt f_i \psi_i, \qquad f_i \in \{1,-1\},\ee 
 where $\psi_i$ is the Ising field on layer $i$. 
 
 We will address this question within the context of the $\ep$ expansion. 
 Our starting point is then the critical action
 \bea \label{philayeraction} S& = \int d^{3}x\, \Big(\,  \frac12 \sum_{i=1}^L(\p\psi_i)^2 
 + \frac{\La}8\sum_{i, j = 1}^L \psi_i^2 \psi_j^2 g_{i,j}\Big),\eea
 where the couplings $g_{i,j}$ are symmetric and dimensionless, $\La$ is a UV cutoff, and where the absence of mass terms denotes tuning to the critical point.  
 
 Enumerating all the symmetric RG fixed points for even moderately small $L$ is essentially impossible, since the symmetry group $\zt^L$ possesses many quartic invariants. 
 However, since we are only interested in symmetric fixed points which are {\it stable} (viz. those which have no relevant quartic terms), the situation becomes much more tractable. This is because we may take advantage of powerful results about the properties of 1-loop beta functions, which follow from interpreting the RG flow as a gradient flow on the space of quartic couplings.
 
 In the following we will make use of two facts. The first is that for any subgroup $H \subset O(L)$ (we will be concerned with $H = \zt^L$), there is {\it at most one} stable $H$-symmetric fixed point.\cite{michel1984renormalization,zinn2002quantum} The second is that the action of $O(L)$ on the $g_{i,j}$ couplings maps fixed points to fixed points, since for the purposes of calculating the beta functions $O(L)$ transformations are simply redundant relabelings of the fields. 
 
 These two facts mean that if the $g_{i,j}$ couplings describe a stable symmetric fixed point, the $g_{i,j}$ must be invariant under the action of any $O(L)$ transformation which preserves the $\zt^L$ symmetry (otherwise the fixed point would not be unique). In particular, consider the action of $\sigma\in S_L \subset O(L)$ on a given $\zt^L$-invariant coupling $g_{i,j}$. The action of $\s$ takes 
 $g_{i,j} \mt g_{\sigma(i),\sigma(j)}$, and therefore a necessary condition for $g_{i,j}$ to give a $\zt^L$-stable fixed point is for $g_{\sigma(i),\sigma(j)} = g_{i,j}$ for all $\sigma\in S_L$. This means that any symmetric stable fixed point will have couplings of the form 
 \be \label{ztlstable} g_{i,j}^* = (g_0-h)\delta_{i,j} + h\ee 
 for some constants $g_0, h$. If $h\neq 0$ this set of couplings gives very non-local interactions, as it contains all-to-all quartic interlayer couplings. On physical grounds we may then be justified in restricting our attention to $h=0$. However, even in the general case with nonzero $h$, we will see that no choice of $g_0,h$ gives a stable fixed point. 
 
 Indeed, it is not hard to explicitly compute the  one-loop $\beta$ functions at fixed points of the form \eqref{ztlstable} and show that as long as $L>4$, none of them are stable (at least within the context of the $\ep$ expansion). We find 
 \bea \beta_{g_0}& = g_0 - \frac92 g_0^2 - \frac{L-1}2 h^2 \\  \beta_h & = h - 3g_0h - \frac{L+2}2 h^2.\eea
There are three fixed points to these equations. One is the decoupled fixed point where $h=0$, which is unstable. The other two are the $O(N)$ symmetric fixed point $\mcs$ and the cubic fixed point $\mcc$. The couplings at each are given by 
 \be (g_0^*, h^*) = \begin{dcases} \mcc \, : \,  & \( \frac{2}9 (1-1/L), \, \frac{2}{3L}\) \\ \mcs \, : \, & \( \frac{2}{8+L},\, \frac{2}{8+L}\) \end{dcases}\ee 
 It is then straightforward to check that both fixed points are unstable (provided that $L>4$).
 
 The above discussion has focused on the RG flows obtained in a perturbative expansion about the free fixed point. 
 One might imagine a possible way out by first introducing a strong deformation that drives the system to a different fixed point, around which the RG analysis is modified. The simplest possibility is to add the term  
 \be \label{layer_hybridizing} \d S = \sum_{i} g_{pair} \int d^2x\, d\tau\, \ep_{2i} \ep_{2i+1}.\ee  
 In the absence of further couplings between pairs of layers, this term drives the system to a decoupled stack of 2d XY models.\footnote{Indeed, two 2d Ising CFTs coupled by their energy operators flow in the IR to the 2d XY fixed point, which can be derived either from the $\ep$ expansion or from conformal perturbation theory, using the known values\cite{poland2019conformal} for the OPE coefficients at the Ising fixed point. } 
 
 This fixed point is however also unstable. Indeed, as was mentioned in the main text, the energy operator $\ep_{i,XY}\sim \ep_{2i} + \ep_{2i+1}$ in the 2d XY model is known to have scaling dimension $\Delta_\ep \approx 1.51>3/2$,\cite{poland2019conformal} and as such energy-energy couplings between the XY layers are irrelevant. However, the $\zt^L$-invariant operators $t_i^{XY} = \ep_{2i} - \ep_{2i+1}$ (which in a given XY layer is $\sim\phi^2 + (\phi^*)^2$) has dimension $\De_t \approx 1.24<3/2$,\cite{chester2020carving} and hence couplings between $t_i$ operators on neighboring layers are relevant, destabilizing the fixed point. 
 
 Therefore while we do not have a rigorous proof, the phase transition that occurs in the $N=2$ case generically seems to be driven first-order by fluctuations, as was suggested by the numerics of Ref. \onlinecite{slagle2017fracton}.   

 \section{Effects of the ring-exchange term \label{app:ring_exchange}} 
 
 In this appendix we discuss what happens to the critical point \eqref{free_chains} in the limit where the chains are decoupled and the ring-exchange term is the dominant relevant perturbation (as discussed in section \ref{sec:oned_critical_largen}, for $N>4$ there is always a region where the ring-exchange term is the only relevant perturbation to the decoupled fixed point). 
  
 When the ring-exchange term is relevant, we are prompted to expand the cosine as $1-\frac12 a^2(\p_x\p_y \cp)^2$, where now the field $\cp$ fluctuates in all four spacetime directions. Doing this and integrating out the $\ct_w$ fields\footnote{The vertex operators $e^{i\ct_w}$ create vortices in $\cp$ in the $z$-$\tau$ plane, which are very energetically costly in the presence of gradient terms in the $x$ and $y$ directions. As such it is best to first integrate them out and then work entirely in terms of the $\cp$ fields.} then gives the continuum action
 \bea S & = \l \int d^3x\, d\tau\, \( a^{-2} [(\p_\tau \cp)^2 +  (\p_z\cp)^2] + \a (\p_x\p_y \cp)^2 \),\eea 
 where $\cp$ and $\a$ are dimensionless and $\l$ is some non-universal parameter. This is a variant of the Bose plaquette model appearing in the analysis of exciton Bose liquids,\cite{paramekanti2002ring,you2020fracton,seiberg2020exoticI,seiberg2020exoticII} which differs from the original model by the presence of an extra spatial dimension and the $(\p_z \cp)^2$ term. In the original model, the degeneracy of the dispersion along the $k_x = 0$ and $k_y = 0$ axes in momentum space leads to strong IR divergences which prevents ordering, leading to a stable massless phase. As we will see, things are rather different in the present case.
 
 The symmetry-invariant operators we may consider correlation functions of are polynomials in $\p_z\cp , \p_\tau\cp ,$ and $\p_x\p_y\cp$, as well as exponentials of $N\cp$. 
 On one hand, exponentials of single $N\cp$ operators have correlation functions which essentially vanish. Indeed, looking at correlation functions along the $\tau$ direction and setting $\a = 1$ for simplicity, we calculate
 \bea \label{anebl_corr} \lan \cp(\tau)\cp(0)\ran & = \frac{\twp a^2}{R^2} \int_{\bfk,\o} \frac{e^{i\tau \o}}{\o^2 + k_z^2 + (ak_xk_y)^2 }\\ 
 & = \frac{a^2}{2R^2} \int_{k_x,k_y} \int_{a|k_xk_y|}^\infty du  \frac{e^{-\tau u}}{\sqrt{u^2 - (ak_xk_y)^2}} \\ & = 
 \frac{a^2}{2 \pi^2 R^2} \int dk_x\, dk_y\, K_0(ak_xk_y \tau )
 \\ & \sim \frac{a}{\fpi R^2 \tau} \ln(L/a). \eea 
 where $L$ is an IR cutoff, with $L/a \ra \infty$ in the thermodynamic limit. As such the correlator 
 \be \lan e^{i N \cp(\tau) } e^{-iN\cp(0)} \ran \sim \( \frac{a}{L}\) ^{\varsigma (1 - a/\tau)} \ra 0,\ee
 where $\varsigma = N^2 /\fpi R^2$, is ultra short-ranged. 
 
 However, correlation functions of exponentials which create $N$ lineon dipoles on neighboring wires (which preserve the subsystem symmetry due to the factor of $N$) do not vanish in this way. In the continuum theory these operators map to exponentials of $N\p_{x}\Phi$ or $N \p_y \Phi$, and the derivatives are able to eliminate the IR divergence encountered in the momentum integration of \eqref{anebl_corr}. For example, by a similar calculation as above we find 
 \bea \lan &e^{i N[ \cp( \tau,x +a) - \cp(\tau,x)]} e^{-iN[\cp(0,x+a)-\cp(0,x)]} \ran \\  &\qquad  \sim \lan e^{iNa\p_x \cp(\tau)} e^{-iNa\p_x\cp(0)}\ran \\  & \qquad \sim \exp\( \frac{a^3 N^2}{2\pi^2 \tau R^2} \int_{1/L}^{1/a} dk_x\, k_x \) \\ & \qquad \sim 
 \exp\( \frac{aN^2}{4\pi^2 \tau R^2}\), \eea 
 which goes to a non-zero value in the limit $\tau/a\ra\infty$ (the same behavior occurs for correlation functions along different spacetime directions). The dispersion along the $k_z$ direction is therefore enough to soften the IR divergences {coming from unusual dispersion in $k_x$ and $k_y$}, enabling the model to order in the IR.
 
 Indeed, upon adding the subsystem-symmetry-allowed terms $\cos(N a\p_{x,y} \Phi)$ to the action, the above correlation function means that lineon dipoles condense. Fluctuations about the condensate give rise to $(\partial_{x}\Phi)^2,(\p_y\Phi)^2$ terms in the action, which eliminates the IR divergences that prevent $\Phi$ from fully condensing. 
 {Therefore the relevance of the ring-exchange term does not induce a flow to an intermediate massless phase,} so that the transition into the condensed phase is expected to be first-order. 
 
 While all of the above discussion has been within the context of condensing a single species of lineon, the case when multiple species condense can be treated in the same way, as the most relevant interwire couplings continue to be those dealt with above.

\bibliography{subdimensional_criticality}

\end{document}

%% file: subdimensional_criticality_preamble.tex
\usepackage{graphicx}
\usepackage[nointegrals]{wasysym} 
\usepackage[export]{adjustbox}
\usepackage{amsmath,amsfonts,amssymb,latexsym}
\usepackage{hhline}
\usepackage{bm}
\usepackage{verbatim}
\usepackage{enumitem}
\usepackage{mathrsfs}
\usepackage{slashed}
\usepackage{empheq}

\newcommand{\p}{\partial}

\newcommand{\lan}{\langle}
\newcommand{\ran}{\rangle}

\newcommand{\unit}{\mathbf{1}}

\newcommand{\La}{\Lambda}

\newcommand{\da}{{\dagger}}

\newcommand{\ra}{\rightarrow}
\newcommand{\lra}{\leftrightarrow}

\newcommand{\wt}{\widetilde}

\newcommand{\uva}{{\mathbf{\hat a}}}
\newcommand{\uvb}{{\mathbf{\hat b}}}
\newcommand{\uvc}{{\mathbf{\hat c}}}

\newcommand{\uvx}{{\mathbf{\hat x}}}
\newcommand{\uvy}{{\mathbf{\hat y}}}
\newcommand{\uvz}{{\mathbf{\hat z}}}

\newcommand{\bfzero}{{\mathbf{0}}}

\renewcommand{\(}{\left(}
\renewcommand{\)}{\right)}

\newcommand{\mt}{\mapsto}

\newcommand{\tp}{\otimes}

\newcommand{\twp}{{2\pi}}
\newcommand{\fpi}{{4\pi}}

\newcommand{\D}{\nabla}

\newcommand\bpm            {\begin{pmatrix}}
	\newcommand\epm           {\end{pmatrix}}

\def\app#1#2{%
	\mathrel{%
		\setbox0=\hbox{$#1\sim$}%
		\setbox2=\hbox{%
			\rlap{\hbox{$#1\propto$}}%
			\lower1.1\ht0\box0%
		}%
		\raise0.25\ht2\box2%
	}%
}

\newcommand{\cp}{\Phi}
\newcommand{\ct}{\Theta}

\newcommand{\ope}\odot

\usepackage{manfnt}

\renewcommand{\prl}{{\, \parallel \,}}
\newcommand{\bi}{\begin{itemize}}
\newcommand{\ei}{\end{itemize}}

\usepackage{amsthm}

\newtheorem{theorem}{Theorem}

\theoremstyle{definition}
\newtheorem{definition}{Definition}
\theoremstyle{definition}

\newcommand\bd            {\begin{definition}}
\newcommand\ed            {\end{definition}}
\newcommand\bt            {\begin{theorem}}
\newcommand\et            {\end{theorem}}
\newcommand\be            {\begin{equation}}
\newcommand\ee            {\end{equation}}
\newcommand\ba            {\begin{aligned}}
\newcommand\ea            {\end{aligned}}
\newcommand\bea{\begin{equation}\begin{aligned}}
\newcommand\eea{\end{aligned}\end{equation}}

\usepackage{subcaption}

 \usepackage{hyperref} 
 \hypersetup{final}
 \hypersetup{colorlinks, citecolor=red, linkcolor=red, urlcolor=red} 
 
 \newcommand{\sss}{\subsubsection}
 \renewcommand{\ss}{\subsection}
 
 \renewcommand{\a}{\alpha}
 \renewcommand{\b}{\beta}
 \renewcommand{\d}{\delta}
 \newcommand{\De}{\Delta}

 \newcommand{\s}{\sigma}

 \newcommand{\ep}{\varepsilon} 
 \renewcommand{\l}{\lambda}
 \renewcommand{\L}{\Lambda}

 \renewcommand{\o}{\omega}

 \newcommand{\mfd}{\mathfrak{d}}
 \newcommand{\mfe}{\mathfrak{e}}

 \newcommand{\mfm}{\mathfrak{m}}

 \newcommand{\bfa}{\mathbf{a}}
 \newcommand{\bfb}{\mathbf{b}}

 \newcommand{\bfk}{\mathbf{k}}

 \newcommand{\bfq}{\mathbf{q}}

 \newcommand{\zt}{\mathbb{Z}_2}
 \newcommand{\zn}{\mathbb{Z}_N}

 \newcommand{\zz}{\mathbb{Z}}
 
 \newcommand{\mcc}{\mathcal{C}}
 
 \newcommand{\mcb}{\mathcal{B}}

 \newcommand{\mcu}{\mathcal{U}}

 \newcommand{\mcs}{\mathcal{S}}

 \newcommand{\mcx}{\mathcal{X}}
 \newcommand{\mcz}{\mathcal{Z}}
 \newcommand{\mcm}{\mathcal{M}}

 \newcommand{\sfc}{\mathsf{c}}

 \newcommand{\sfi}{\mathsf{i}}
 \newcommand{\sfj}{\mathsf{j}}
 
 \newcommand{\sfl}{\mathsf{l}}

 \newcommand{\sfp}{\mathsf{p}}

 \newcommand{\sfx}{\mathsf{x}}
 
 \newcommand{\sfz}{\mathsf{z}}
 
 \usepackage[mathscr]{eucal} 

 \newcommand{\sq}{\square}
 
 \newcommand{\tA}{\widetilde{A}}
 \newcommand{\tZ}{\widetilde{Z}}
 \newcommand{\tX}{\widetilde{X}}

\usepackage{braket}